\documentclass[aps,preprintnumbers]{revtex4}
\usepackage[T1]{fontenc}
\usepackage[utf8]{inputenc}
\usepackage{graphicx}
\usepackage{epsfig}
\usepackage{ulem}
\graphicspath{{./Figures/}}                 % location for color  fig.
\usepackage[colorlinks=true,linktocpage=true,linkcolor=blue,citecolor=blue,allcolors=blue]{hyperref}
\usepackage{amsmath}
\usepackage{amssymb}

\usepackage{array}
\usepackage{picture}
\usepackage{multirow}

\newcommand{\beq}{\begin{equation}}
\newcommand{\eeq}{\end{equation}}
\newcommand{\bea}{\begin{eqnarray}}
\newcommand{\eea}{\end{eqnarray}}

\newcommand{\bc}{\begin{center}}
\newcommand{\ec}{\end{center}}

\newcommand{\ds}{\displaystyle}

\newcommand{\acem}{a^{\mathrm{CEM}}_n}
\newcommand{\arfm}{a^{\mathrm{RFM}}_n}

\newcommand*\xbar[1]{%
  \hbox{%
    \vbox{%
      \hrule height 0.5pt % The actual bar
      \kern0.5ex%         % Distance between bar and symbol
      \hbox{%
        \kern-0.1em%      % Shortening on the left side
        \ensuremath{#1}%
        \kern-0.1em%      % Shortening on the right side
      }%
    }%
  }%
}

\usepackage[dvipsnames]{xcolor}
\colorlet{LightRubineRed}{RubineRed!70!}
\colorlet{Mycolor1}{green!10!orange!90!}
\definecolor{Mycolor2}{HTML}{00F9DE}

\begin{document}
\title{Quark Density in Lattice QC$_2$D at Imaginary and Real Chemical Potential}
\author{A.~M.~Begun}
\affiliation{Nordita, Stockholm University, Roslagstullsbacken 23, SE-106 91 Stockholm, Sweden}
\affiliation{Pacific Quantum Center, Far Eastern Federal University, 690950 Vladivostok, Russia}
\author{V.~G.~Bornyakov}
%\email[]{bornvit@gmail.com}
\affiliation{Institute for High Energy Physics of the NRC Kurchatov Institute, 142281 Protvino, Russia}
\affiliation{National Research Centre “Kurchatov Institute”, 123182, Moscow, Russia}
\affiliation{Pacific Quantum Center, Far Eastern Federal University, 690950 Vladivostok, Russia}
\author{ N. V. Gerasimeniuk}
\affiliation{Pacific Quantum Center, Far Eastern Federal University, 690950 Vladivostok, Russia}
\author{V.~A. Goy}
\affiliation{Institut Denis Poisson UMR 7013, Universit\'e de Tours, 37200 Tours, France}
\affiliation{Pacific Quantum Center, Far Eastern Federal University, 690950 Vladivostok, Russia}
\author{A. Nakamura}
\affiliation{RCNP, Osaka University, Osaka 567-0047, Japan}
\affiliation{Pacific Quantum Center, Far Eastern Federal University, 690950 Vladivostok, Russia}
\author{R.~N.~Rogalyov}
\affiliation{Institute for High Energy Physics of the NRC ``Kurchatov Institute'', 142281 Protvino, Russia}
\begin{abstract}

We study lattice two-color QCD (QC$_2$D) with 
two flavors of staggered fermions at imaginary 
and real quark chemical potential $\mu_q$ and $T>T_c$. 
We employ various methods of extrapolation 
of the quark number density from imaginary 
to real quark chemical potentials $\mu_q$,
including series expansions as well as analytic continuation based on phenomenological models,
and study their accuracy by comparing the results 
to the lattice data. Below the Roberge-Weiss temperature, 
$T<T_{RW}$, we find that the cluster expansion model
provides an accurate analytic continuation of the baryon 
number density in the studied range of chemical potentials.
On the other hand, the behavior of the reconstructed 
canonical partition functions indicates that the available models may require corrections at high quark densities.
At $T>T_{RW}$ we 
show that the analytic continuation 
to the real values of $\mu_q$ based on 
trigonometric functions works equally well with 
the conventional method based on 
the Taylor expansion in powers of $\mu_q$.
 
\end{abstract}

\maketitle
\section{Introduction}

Studies of strongly interacting matter under 
extreme conditions are among the most important topics 
in modern high-energy physics. 
Experimentally, the QCD phase structure is probed 
by heavy-ion collisions~(HIC) at the LHC \cite{Aamodt:2008zz}, RHIC \cite{Adams:2005dq}, 
and SPS~\cite{Abgrall:2014xwa}, where a dense and hot fireball 
of QCD matter is created and
then expands and evolves from
the state of strongly-coupled quark-gluon plasma (sQGP) 
to the hadron resonance gas.
In the plane ``baryon chemical potential -- temperature'' 
($\mu_B-T$) these two phases are separated by a transition region.
This region corresponds to an analytic crossover 
at vanishing net-baryon number density,
as established from 
first-principle lattice QCD simulations~\cite{Aoki:2006we}.
It is feasible that the transition line turns into 
a first-order phase transition at high net-baryon number densities with an associated QCD critical end point (CEP).
Analysis of the QCD phase structure 
at finite baryon number densities is in the focal point 
of the  beam energy scans at RHIC~\cite{Bzdak:2019pkr,Adam:2020unf} and SPS~\cite{Gazdzicki:2015ska}, which will be supplemented 
by the future HIC experiments at FAIR~(GSI) \cite{Ablyazimov:2017guv} 
and at NICA~(JINR) \cite{Sissakian:2009zza}.
It is hoped that these experiments will be able 
to answer the question of the existence of the CEP. 

Presently we have no proper theoretical understanding 
of the phase diagram of QCD in the $\mu_B-T$ plane 
from first principles. In particular, even the dependence 
of the net-baryon number density $n_B$ 
on the quark chemical potential 
$\mu_q \equiv \mu_B / N_c$ remains an open question, 
though we know that
$n_B$ increases with increasing $\mu_q$ and $n_B(\mu_q=0)=0$.
While at $\mu_q=0$ the QCD phase diagram has  
been successfully  studied theoretically 
in the framework of lattice QCD, 
at finite $\mu_q$ in the range relevant for the CEP search 
this approach is plagued by the sign problem. 

Even though direct lattice simulations at $\mu_q\neq 0$ are currently
not possible, the thermodynamical quantities can be calculated 
at small $\theta \equiv \mu_q/T$ indirectly. 
Two approaches have commonly been employed: the first one is based on  
Taylor expansion in $\theta$ around $\theta = 0$~\cite{Bellwied:2015lba,Ding:2015fca,Bazavov:2018mes,Bazavov:2020bjn} while the second one employs analytic continuation from imaginary $\theta$, where sign problem is absent, to real $\theta$~\cite{DElia:2002tig,DElia:2009pdy,Bonati:2014kpa,Takahashi:2014rta,DElia:2016jqh,Bornyakov:2016wld,Alba:2017mqu,Bornyakov:2017upg,Bonati:2018nut,Borsanyi:2018grb}.

In this work we study two-color QCD (QC$_2$D) on the lattice, 
which have received considerable attention in the literature, 
see, e.g. \cite{Nakamura:1984uz,Hands:1999md,Kogut:2001if,Kogut:2002cm,Muroya:2002ry,Giudice:2004se,Hands:2006ve,Cea:2006yd,Cea:2007vt,Cea:2009ba,Cotter:2012mb,Boz:2013rca,Braguta:2016cpw,Holicki:2017psk,Bornyakov:2017txe,Boz:2018crd,Astrakhantsev:2018uzd,Boz:2019enj,Iida:2019rah,Wilhelm:2019fvp,Bornyakov:2020kyz,Astrakhantsev:2020tdl,Buividovich:2020dks,Iida:2020emi} and references therein. 
The keen interest in QC$_2$D and QCD-like theories 
stems from two reasons: first, they are expected 
to share common properties with full QCD in some 
parts of their phase diagrams, and, 
second, they are numerically tractable, 
thus allowing to test methods that can later be used in QCD. 
In particular, the lattice simulations of QC$_2$D 
can be performed both at imaginary and real $\mu_q$.
We use this fact to analyze the efficiency 
of various procedures based on the analytic-continuation 
method and select the optimal one. 
The results of earlier studies along these lines were presented in Refs.~\cite{Giudice:2004se,Cea:2006yd,Cea:2007vt,Cea:2009ba}. 

The paper is organized as follows.
In Sec.~\ref{sec:method} we describe details 
of our lattice simulations. Secs.~\ref{sec:lowT} 
and~\ref{sec:highT} explore the performance of various  procedures of extrapolation from imaginary to real chemical potential at low and high temperatures, respectively.
Summary in Sec.~\ref{sec:conclusions} closes the article.

\section{Definitions and details of simulation}
\label{sec:method}

The grand canonical partition function  $\ds Z_{GC}\left({\frac{\mu_q}{T}} \equiv \theta, T, V\right)$ can be written as the sum of the canonical ones $Z_C(n, T, V)$: 
\begin{eqnarray} \label{ZG}	
Z_{GC}(\theta, T, V)=\sum_{n=-\infty}^\infty Z_C(n,T,V)\xi^n,
\quad
\end{eqnarray}
where
$\xi=e^{\theta}$ is the fugacity. For brevity, we omit $T$ and $V$ from the arguments of $Z_{GC}$ and $Z_C$. 
We refer to the expression given by Eq.~(\ref{ZG}) 
as the fugacity expansion. Here and below, the variable 
$\ds \theta = {\mu_q\over T} = \theta_R +\imath \theta_I$ 
is employed.

The fugacity expansion represents the Laurent series 
in powers of $\xi$,
with coefficients $Z_C(n)$ which are uniquely determined 
only in the case when $Z_{GC}(\theta)$  
is an entire function of $\xi$, which takes place for systems 
in a finite volume.
However, in the present study we are confronted 
with some models where this is not the case and $Z_{GC}(\theta)$
has singularities in the complex $\xi$ plane.
In these situations, we focus our attention on the annulus including the circle $|\xi|=1$ in which $Z_{GC}(\theta)$ is analytic. Within this annulus, the coefficients $Z_C(n)$ in Eq.~(\ref{ZG}) coincide with those for the Fourier expansion 
\beq \label{eq:fug_exp_unit_circle}
Z_{GC}(\imath \theta_I)=\sum_{n=-\infty}^\infty Z_C(n)\exp(\imath n  \theta_I )\;, 
\eeq 
whereas, beyond this annulus, the coefficients $Z_C(n)$ 
in  Eq.~(\ref{ZG}) differ from those in Eq.~(\ref{eq:fug_exp_unit_circle}).
The inverse of the expansion (\ref{eq:fug_exp_unit_circle}) has the form
\cite{Hasenfratz:1991ax}
\begin{eqnarray} \label{Fourier}
Z_C\left(n\right)=\int_{0}^{2\pi}\frac{d\theta_I}{2\pi}
e^{-in\theta_I}Z_{GC}( \imath \theta_I )\; .
\end{eqnarray}
$Z_{GC}$ is a periodic function
of $\theta$: $Z_{GC}(\theta) = Z_{GC}(\theta+2\pi\imath /N_c)$.
As a consequence of this periodicity the canonical partition functions $Z_C(n)$ are nonzero only for $n=N_c \cdot k$ for $k \in \mathbb{Z}$. This symmetry is called the Roberge-Weiss symmetry \cite{Roberge:1986mm}. 
It should be emphasized that $Z_C(n)$ are proportional to the probability mass function for the distribution in the net-baryon number and, therefore, they are of phenomenological significance
\cite{Nakamura:2013ska}.

Below we will mostly use the following 
dimensionless variable representing 
the net-baryon number in the lattice 
volume $V=a^3 N_s^3$ under consideration
\beq
B={n_q V \over N_c}
\eeq
instead of the quark number density $n_q$. These quantities 
are related by a factor that does not change 
throughout our study. The net-baryon number $B$ 
for $N_f$ degenerate quark flavors is determined 
by the expression
\bea\label{eq:n_Z}
B(\theta) &=& \frac{1}{N_c}\frac{\partial\;\ln Z_{GC}}{\partial \theta}  \nonumber  \\
     &=&  \frac{N_{f}}{N_c Z_{GC}} \int \mathcal{D}U e^{-S_G} 
     [\det\Delta(\theta)]^{N_f}
\mathrm{tr}\left[\Delta^{-1}\frac{\partial \Delta}{\partial \theta}\right]\;, 
\eea
where $S_G$ is the lattice gluon action, $\Delta$ 
is the lattice Dirac operator. It is clear that if $Z_{GC}(\theta)$ 
is an entire function then $B(\theta)$ is meromorphic.

We compute $B(\theta)$ numerically in QC$_2$D 
at both imaginary and real $\mu_q$.
From Eqs.~(\ref{ZG}) and (\ref{eq:n_Z}) it follows 
that $B(\theta)$ can be expressed in terms 
of the normalized canonical partition functions
$Z_n=Z_C(nN_c) / Z_C(0)$  as follows:
\beq
\label{density_re}
B(\theta) = \frac{2\sum_{n=1}^\infty n Z_n \sinh(nN_c\theta)}{1+2\sum_{n=1}^\infty Z_n \cosh(nN_c\theta)}~~~,
\eeq
and, in particular,
\beq
\label{density2}
B(\theta)  = \frac{2\imath\; \sum_{n=1}^\infty n Z_n \sin(nN_c\theta_I)}{1+2\sum_{n=1}^\infty Z_n \cos(nN_c\theta_I)}\qquad \mbox{for} \qquad \theta_R=0\;.
\eeq
Our calculations correspond to $N_c=2$.
Where applicable, we do provide the expressions for the general case of arbitrary $N_c$.
Note that $Z_n$ define the probability
${\cal Z}(n)$ that a system sampled from the grand-canonical ensemble at $\mu_q=0$ has the baryon number $n$, namely $\ds {\cal Z}(n)={Z_{|n|}\over 1+2\sum_{k=1}^\infty Z_k}$. 

The details of our lattice setup are as follows.
We employ the tree level improved Symanzik gauge action~\cite{Weisz:1982zw} and staggered fermion action. We do not include the diquark source~\cite{Hands:1999md} since at temperatures considered in this paper the diquark condensate is zero \cite{Cotter:2012mb}.
More details about our lattice action 
can be found  in~\cite{Bornyakov:2017txe}.
We perform simulations on $N_s^3 \times N_t$ lattices 
at $\beta=1.7$ and fix the scale
using the Sommer parameter value $r_0=0.468$~fm. 
The corresponding lattice spacing is approximately $0.062$~fm.
We consider $N_s=28$ which gives lattice size 
$L\approx 1.74$~fm, and the set of temperatures 
$T=227, 265, 398$ and 530~MeV corresponding 
to $N_t=14,12,8$ and~6, respectively.    
We employ the quark mass value in lattice units $am_q=0.0125$, the respective pion mass is rather large, $m_{\pi} \approx 800$~MeV. 
The simulations are performed at imaginary quark 
chemical potential over the range $0< \theta_I < \pi/2$  
with statistics between 2000 and 5000 configurations
and at real quark chemical potential over the range 
$0< \mu_q  \lesssim 600$~MeV  with statistics between 1000 and 3000 configurations. For temperature $T=265$~MeV statistics was increased at imaginary $\mu_q$ up to 24000 configurations. 

Back in 1986 Roberge and Weiss argued \cite{Roberge:1986mm} that 
there exists the temperature $T_{RW}$ such that at $T<T_{RW}$ 
the quark number density is a smooth function of $\theta_I$, 
whereas at $T>T_{RW}$
it has discontinuities at $\ds \theta_I=\pi(2 n +1)/ N_c,~n\in Z\!\!\!Z$. 
In $N_f=2+1$ lattice QCD at physical quark masses 
the value  $T_{RW}=208(5)$~MeV~\cite{Bonati:2016pwz} 
was found, i.e. $T_{RW}/T_{pc} = 1.34$ in that theory. In our study of QC$_2$D we found a smooth dependence 
of the baryon density on $\theta_I$ at $T=265$~MeV, whereas 
at $T=398$~MeV it is discontinuous at $\ds \theta_I=\pi/2$. 
Thus, we conclude that 265 MeV $< T_{RW} <$ 398 MeV. 
Our preliminary results at $N_t=10$ (not presented 
in this paper) indicate that  $T_{RW} \sim 320$ MeV.
In this study we explore the cases $T<T_{RW}$ 
and $T>T_{RW}$ separately. The former case is referred to 
as ``low temperatures'' and the latter---``high temperatures''.

\section{Analytic continuation of the baryon number density: 
low temperatures}
\label{sec:lowT}

In real QCD for temperatures $T<T_{RW}$ it is common \cite{DElia:2009pdy,Takahashi:2014rta,Bornyakov:2016wld} to 
perform the extrapolation from imaginary to real values of $\theta$ by employing the expression for the net-baryon number in the form of a trigonometric Fourier series,
\beq\label{eq:Fourier_expansion}
\tilde B^{(a)}_N(\theta_I)=\sum_{n=1}^N a_n \sin\left({2 n \theta_I}\right)~.  \\ 
\eeq
By fitting the lattice data over the segment 
$\ds 0\leq \theta_I \leq {\pi\over 2} $ one first determines 
the coefficients $a_n$ of the expansion and then 
considers $\imath \tilde  B^{(a)}_N(\theta_I)$ 
as the analytic function $B^{(a)}_N(\theta)$ 
at $\theta=\imath \theta_I$. 
On the real axis of the complex $\theta$ plane it has the form
\beq 
B^{(a)}_N(\theta)\Big|_{\theta_I=0} = \sum_{n=1}^N a_n \sinh\left({2 n \theta_R}\right) \;, 
\label{eq:naive_anal_cont}
\eeq
and describes the baryon number at real values of the chemical potential. 
However, there exists a problem of convergence in the limit
$n\to\infty$. More precisely,  if the limit 
\beq 
\lim_{n\to\infty} {|a_{n+1}|\over |a_n|} = j 
\eeq 
exists and $0<j<1$, then the series 
(\ref{eq:naive_anal_cont}) converges in a finite region
\beq 
|\theta_R|< {-\; \ln j\over 2}\ . 
\label{eq:diverg_of_sinhs}
\eeq 
$j=1$ corresponds to vanishing radius of convergence of the series 
(\ref{eq:naive_anal_cont}) while for
$j \to 0$ it tends to infinity. In practice, the radius of convergence may 
be rather small, and in such a case 
this  procedure loses practical relevance. 
In subsection \ref{subsec:anal_cont} we illustrate this using a model.
This motivates one to search for a different method of extrapolation. 

\subsection{Determination of $Z_n$ from a direct fit to data}

One natural method for $T<T_{RW}$ is to use expressions 
(\ref{density_re}) and (\ref{density2}).
The series in both the numerator and the denominator 
of these expressions converge because $Z_n$ decrease
more rapidly than the geometric progression~(see~\cite{Bornyakov:2016wld} for an explicit demonstration).
In this case, the right-hand sides  
in Eqs.~(\ref{density_re}) and (\ref{density2})
define the density in the entire complex plane of $\theta$.
The difficulty lies in the evaluation of a sufficient number 
of $Z_n$ using a limited data set for the quark density. 
As shown below, our data allows to compute 
only very few of the leading $Z_n$.

We fit our data at imaginary chemical potential to the function
\beq
\tilde{B}^{(Z)}_{N}(\theta_I) =
 {2 \sum_{n=1}^{N} n Z_n \sin(2n\theta_I) \over 1 + 2 \sum_{n=1}^{N} Z_n \cos(2n\theta_I)}, \qquad \theta_I \in \left[0, {\pi\over 2}\right],
\label{eq:ff_for_Z}
\eeq
inspired by the expansion (\ref{density2}).
We find that determination of $Z_n$ via the fit 
by Eq.~(\ref{eq:ff_for_Z}) is not a straightforward matter. We explain the problem and our approach to its solution in Appendix \ref{sec_App1}.

Various values of $N$ have been utilized 
in the fitting procedure employing the method 
of least squares based on minimizing the sum of the squares 
of the residuals under the assumption that these residuals 
are normally distributed
(so that this sum is $\chi^2$-distributed). 
The fit results and the respective parameters characterizing
goodness of fit (both $\ds {\chi^2/N_{dof}}$ and $p$-value) 
are presented in Tables~\ref{tab:direct_fit_14} 
(for $T=227$~MeV) and \ref{tab:direct_fit_12} 
($T=265$~MeV) of Appendix \ref{sec_App2}. 
The results of the fits for the optimal value 
of parameter $N$ are also shown in 
Tables~\ref{tab:compare_Z_n_a_T14} and~\ref{tab:compare_Z_n_a}  
to compare them with the results following from the models. 
These results are also used to produce the fitting curves in Fig.~\ref{fig:CEM_vs_lattice2} (left panel).

We found that the fits based on Eq.~(\ref{eq:ff_for_Z}) produce 
reasonable results for large enough $N$ only, namely, $N \geq 3$ 
for  $T=227$~MeV and $N \geq 8$ for $T=265$~MeV.  
The denominator in Eq.~(\ref{eq:ff_for_Z}) has zeros 
in the complex $\theta$ plane corresponding 
to the Lee-Yang zeros in the $\xi$ plane. 
For small $N$ they appear 
close to imaginary axis and the fit function (\ref{eq:ff_for_Z}) 
demonstrates strong oscillations.
A similar behavior associated with the Lee-Yang zeroes 
has been observed in the earlier studies of Ref.~\cite{Wakayama:2018wkc}.
As for the upper value of $N$, the direct fits for $N>7$ 
at $T=227$~MeV and for $N>9$  at $T=265$~MeV 
are hindered by insufficient statistics.

After $Z_n$ are determined either by the direct fit 
described above or using some model (see below), 
one can employ Eq.~(\ref{eq:ff_for_Z}) for analytic 
continuation. We will call this approach to extrapolation 
the canonical approach. The results of this analytic continuation are shown in Fig.~\ref{fig:CEM_vs_lattice2} 
and will be discussed in Section \ref{subsec:anal_cont}. 
It should be emphasized 
that the method based on the canonical approach is 
model independent and stems from first principles.

\subsection{Model-dependent fit functions}

Another way of extrapolation is to perform the analytic continuation by using model-dependent fit functions.
The advantage is that in such a case one only has to fix a few model parameters, which is feasible to do using the available lattice data.
One can then calculate as many coefficients $a_n$ as needed to determine both the $Z_n$ and the behavior of the baryon number at real chemical potentials.
The disadvantage here is the necessity to rely on model assumptions.

We consider two models which have recently been discussed 
in the literature in the context of full QCD: 
the cluster expansion model (CEM)~\cite{Vovchenko:2017gkg} 
and the rational  fraction 
model (RFM) \cite{Almasi:2018lok}, each containing two free parameters.

Let us consider the CEM first. We will use the notation
$\acem$ for the Fourier coefficients $a_n$
in Eq.~(\ref{eq:Fourier_expansion}) computed 
in the framework of the CEM.
The CEM fixes all higher-order Fourier coefficients $\acem$ for $n \geq 3$ as function of the leading two. 
In Ref.~\cite{Vovchenko:2017gkg} the CEM was 
formulated for $N_c = 3$ QCD with 2+1 flavors, 
and was applied to imaginary $\mu_B$ lattice QCD
data~\cite{Vovchenko:2017xad} at physical quark masses.
Here we adopt the CEM for QC$_2$D with two flavors, 
as appropriate for our lattice simulations.
The CEM coefficients can be represented in terms 
of two free parameters $b$ and $q$~(instead of $b_1$ and $b_2$
used in Refs.~\cite{Vovchenko:2017xad,Vovchenko:2017gkg}), which for the two-flavor QCD read as:
\beq
\acem\;=\;(-1)^{n+1}\;{b\, q^{n-1}\over n} \left[ 1+{6\over \pi^2 (N_c^2-1) n^2}  \right] \; ,
\label{eq:CEM_coeffs}
\eeq
where
\beq
b={b_1\pi^2 (N_c^2-1)\over 6 + (N_c^2-1)\pi^2}, \qquad q= -\,{4\,(6 + \pi^2 (N_c^2-1))\over 3+2\pi^2 (N_c^2-1) }\; {b_2\over {b_1}}\;.
\eeq

The RFM is similar to CEM in that it 
also employs the leading two Fourier coefficients 
to fix all other coefficients, and both models match the Stefan-Boltzmann limit of massless quarks. 
However, in contrast to the CEM, where the 
Fourier coefficients exhibit exponential decreasing 
at large $n$, in RFM they obey a power law scaling.
In what follows, the coefficients $a_n$
in Eq.~(\ref{eq:Fourier_expansion}) derived
in the RFM are designated by $\arfm$.
They can be represented in terms of two parameters 
$d$ and $\kappa$ via a concise formula 
\beq\label{eq:RFM_coeffs}
\arfm = (-1)^{n+1} \; d \;{\ds 1+{\pi^2(N_c^2-1)\over 6} n^2 \over n^3 (1 +n\kappa) }\; .
\eeq

We  determine the CEM parameters ($b$ and $q$) and the RFM parameters 
($d$ and $\kappa$)  by fitting summed expressions 
\eqref{CEM_summed} and \eqref{eq:RFM_summed_series_2} 
(see section \ref{subsec:anal_cont}) 
to the lattice data for the baryon number 
over the range $\theta_I \in [0,\pi/2]$. The results of these fits 
are presented in Table~\ref{tab:CEM_fit}. 
The respective statistical errors are evaluated by 
the bootstrap method with a bootstrap sample of size 500.
The $p$ values listed in Table~\ref{tab:CEM_fit} 
indicate that CEM fits our data at imaginary 
chemical potential substantially better than RFM. 
Note that fits to (\ref{eq:Fourier_expansion}) with respective
expressions for $a_n$ and large enough $N$ produce results 
for model parameters in full agreement with those
shown in Table~\ref{tab:CEM_fit}.

%%%%%%%%%%%  BEGIN   TABLE #1    %%%%%%%%%%%%%%%%%
\begin{table}[htb]
\bc
\begin{tabular}{|c|c|c|c|c|c|c|c|c|} \hline
{\multirow{2}{*}{~$T,$~MeV~}} & 
\multicolumn{4}{c|}{CEM} &
\multicolumn{4}{c|}{RFM} \\ 
\cline{2-9}
     &  $p-$value & ~$b$~      &  ~$q$~  & ~$r_{bq}$~  &  $p-$value & ~$d$~      &  ~$\kappa $~ & ~$r_{d\kappa}$~ \\ \hline
%\hline
227  &  0.022  &   1.596(25) & 0.200(31) & 0.700 & 0.004 &  -0.13(16)  &  -1.40(20)  & 0.9993 \\
265  &  0.98  &  4.200(14) & 0.532(7) & 0.785 & $1.5\cdot 10^{-6}$ &  5.1(8)    &   5.0(9) & 1.0    \\
\hline
\end{tabular}
\caption{Parameters $b$ and $q$ of the CEM and $d$ and $\kappa$ of the RFM determined from the fit to our data over the range 
$\ds 0\leq \theta_I \leq {\pi\over 2} $.
The correlations $r_{bq}$ between $b$ and $q$ and $r_{d\kappa}$ 
between $d$ and $\kappa$ as well as respective $p$-values are also shown.
}
\label{tab:CEM_fit}
\ec
\end{table}
%%%%%%%%%%%  END    TABLE #1    %%%%%%%%%%%%%%%%%

The Fourier coefficients $a^{\mathrm{CEM}}_n$ and $a^{\mathrm{RFM}}_n$
related to the CEM and RFM models, respectively, are presented in 
Table~\ref{tab:compare_Z_n_a_T14} (for $T=227$~MeV) 
and in Table~\ref{tab:compare_Z_n_a} (for $T=265$~MeV). 
Their errors were evaluated by the bootstrap method. 
%using bootstrap sample of size 1000.

%%%%%%%%%%%%%%%%%%%%%%%%%%  BEGIN TABLE 2 %%%%%%%%%%%%%%%%%%%%%%%%%%%%%%%
\begin{table}[htb]
\bc
\begin{tabular}{|c|c|c|c|c|c|c|} \hline
                 & & & & & & \\ 
~~n~~  & $a_n$ & ~$\acem$~ &~$\arfm$~&~$Z_n$~& ~$Z_n^{CEM}$~ & ~$Z_n^{RFM}$~ \\ \hline
%\hline
 1  & 1.918(30)   &  1.919(29)   &  1.915(31)   & 0.6702(38) & 0.6686(41)     & 0.6674(42) \\
 2  & -0.139(34)  & -0.168(27)   & -0.185(30)     & 0.2536(37) & 0.2530(46)     & 0.2525(45) \\
 3  & -0.025(37)  &  0.0226(68)  & 0.067(12)   & 0.0567(27) & 0.0621(25)     & 0.0644(24) \\
 4  & -0.048(36)  & -0.0032(16)  & -0.0348(66)    & 0.0060(17) & 0.0107(8)      & 0.01183(79) \\
 5  &  0.067(26)  &  0.00051(36) &  0.0213(41)  & 0.0028(15) & 0.00137(17)    & 0.00181(16) \\
 6  &   ---       & -0.000085(81)& -0.0143(28)  & 0.0017(10) & 0.000135(25)   & 0.000154(35) \\
 7  &   ---       &  0.000015(19)&  0.0103(20) & ---        & 0.0000104(27)  & 0.0000533(13) \\
 8  &   ---       &-0.0000026(44)& -0.0078(15)  & ---        & 0.00000065(23) & - 0.0000217(37)\\
\\
\hline
\end{tabular}
\caption{Fourier coefficients $a_n$ and canonical partition functions $Z_n$ extracted 
from the lattice data at $T=227$~MeV 
by the direct fits (see also Table \ref{tab:direct_fit_14}) as well as with use of the CEM and RFM models.}
\label{tab:compare_Z_n_a_T14}
\ec
\end{table}
%%%%%%%%%%%%%%%%%%%%%%%%%%%%%  END TABLE 2 %%%%%%%%%%%%%%%%%%%%%%%%%%%%%%%

%%%%%%%%%%%%%%%%%%%%%%%%%%%  BEGIN TABLE 3 %%%%%%%%%%%%%%%%%%%%%%%%%%%%%%%%%
\begin{table}[htb]
\bc
\begin{tabular}{|c|c|c|c|c|c|c|} \hline
                 & & & & & & \\ 
~~n~~  & $a_n$ & ~$\acem$~ & ~$\arfm$~  & ~$Z_n$~ & ~$Z_n^{CEM}$~ & ~$Z_n^{RFM}$~ \\ \hline
%\hline
 1  & 5.040 (18) &  5.062(60)   &  5.06(13)  &0.860322(58) & 0.86025(33) & 0.91808(87) \\
2  & -1.175(19) & -1.150(72)   & -1.189(88)  &0.551440(56)& 0.55109(84)   & 0.7108(27) \\
 3  &  0.430(22) &  0.388(45)   &  0.528(50)  &0.26742(17) & 0.26678(91)   & 0.4648(40) \\
 4  & -0.185(25) & -0.150(25)   & -0.298(31)  &0.09997(39)  & 0.09934(60)  & 0.2574(39) \\
 5  &  0.080(23)&  0.062(14)  &  0.191(21)    &0.02938(42)  & 0.02896(28)  & 0.1211(29) \\
 6  &  -0.024(13)& -0.0269(74)  & -0.133(15)  &0.00691(29)  & 0.006719(92)   & 0.0486(17) \\
 7  &   ---      &  0.0120(39)  &  0.098(12)  &0.00127(13)  & 0.001259(24)  & 0.01668(78) \\
 8  &   ---      & -0.0054(21)  & -0.0752(92) &0.000143(28)  & 0.0001930(48)  & 0.00492(30) \\
 9  &   ---      &  0.0025(11)  &  0.0595(74) &  ---         & 0.00002449(76) & 0.001257(96) \\
\hline
\end{tabular}
\caption{Same as in Table \ref{tab:compare_Z_n_a_T14} but for $T=265$~MeV.}
\label{tab:compare_Z_n_a}
\ec
\end{table}
%%%%%%%%%%%%%%%%%%%%%%%%%%%%%  END TABLE 3 %%%%%%%%%%%%%%%%%%%%%%%%%%%%%%%

\subsection{Analytic continuation of the baryon number to real chemical potentials}
\label{subsec:anal_cont}
%%%%%%%%%%%%%%%%% BEGIN FIGURE 1 %%%%%%%%%%%%%%%%%%%%%%%%%%%%
\begin{figure}[htb]
\vspace*{-14mm}
\begin{center}
\hspace*{-4mm}\includegraphics[width=9.0cm]{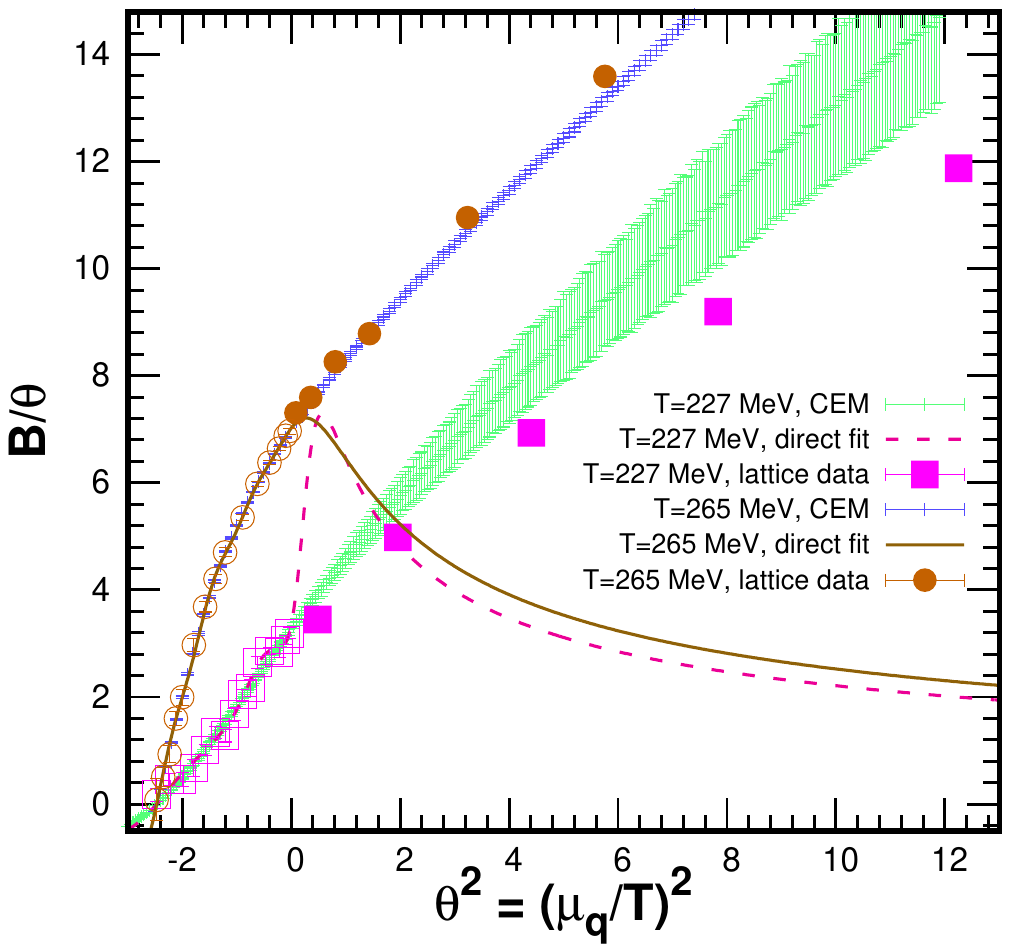}\hspace*{-13mm}\includegraphics[width=9.0cm]{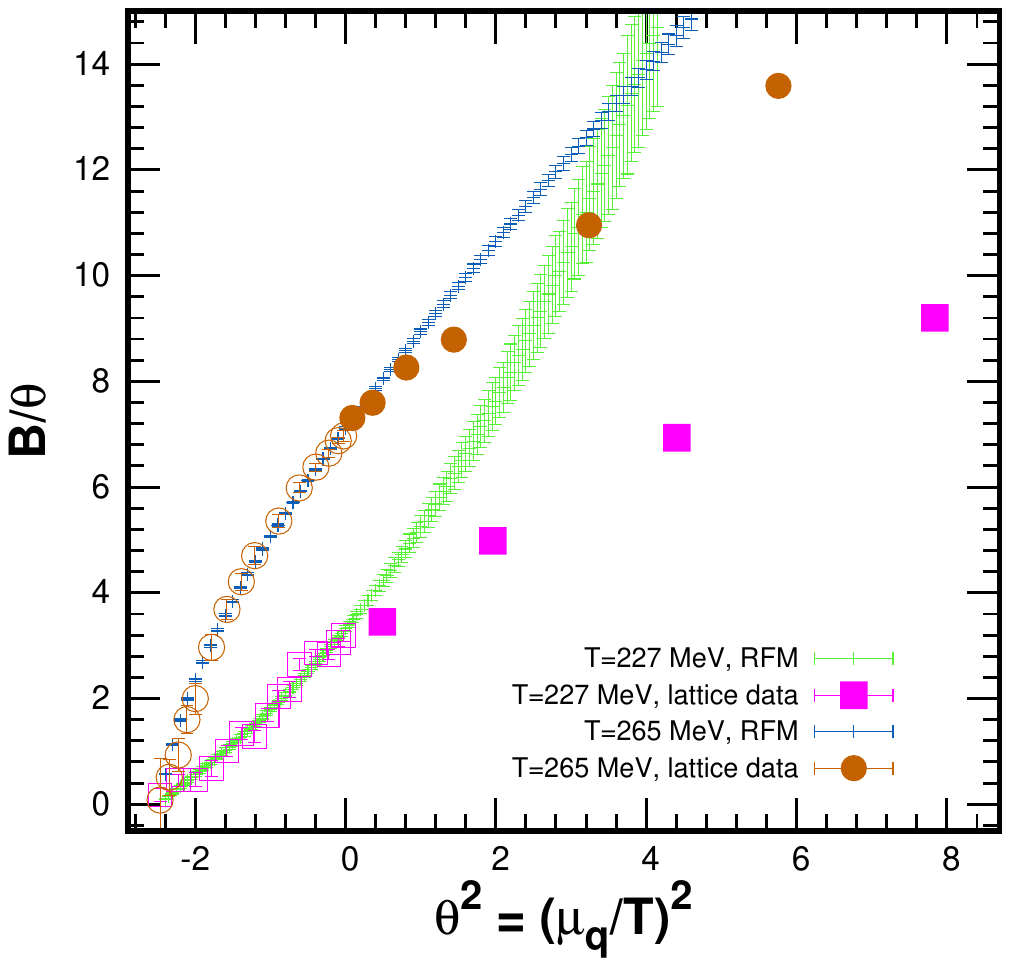}
\vspace*{-18mm}
\caption{The lattice results for the ratio $B/\theta$ as a function of $\theta^2 = (\mu_q/T)^2$ 
are compared with the results of the CEM, Eq.~(\ref{CEM_summed}) (left panel)
and RFM, Eq.~(\ref{eq:RFM_summed_series_2}) (right panel) at temperatures $T=227$~MeV and 265 MeV. The error bands are evaluated by the bootstrap method. The curves in the left panel show results of the fit to Eq.~(\ref{eq:ff_for_Z}) and its extrapolation.
}
\label{fig:CEM_vs_lattice2}
\end{center}
\end{figure}
%%%%%%%%%%%%%%%%% END FIGURE 1 %%%%%%%%%%%%%%%%%%%%%%%%%%%%

On the basis of the above considerations one can naively assume that, at physical values of the quark chemical potential, the expectation value of the baryon number in the lattice volume can be given by the limit 
\beq\label{eq:Naive_limit}
B^{(a)}(\theta_R) = \lim_{N\to\infty} B^{(a)}_N(\theta_R) \equiv \lim_{N\to\infty} \sum_{n=1}^N a_n \sinh\left({2 n \theta_R}\right)\;,
\eeq
where this limit exists. The truncated series $B^{(a)}_N(\theta)$
can be used for analytic continuation provided that the whole series converges. However, the Fourier coefficients $\acem, \arfm$ 
calculated in the models under consideration, 
do not decrease sufficiently fast for the series (\ref{eq:Naive_limit})
to converge in the whole range of relevant chemical potential values. 
The power-like decrease 
of $\arfm$ implies zero radius of convergence of (\ref{eq:Naive_limit}) in $\theta$ plane, therefore, 
the baryon number parameterized in the RFM 
cannot be continued to physical values of 
$\theta$ using partial sums of the series (\ref{eq:Naive_limit}). 
In case of the CEM the series (\ref{eq:Naive_limit})
converges at $\ds |\theta|< -{1 \over 2} \,\ln q$.
Such a small radius of convergence gives rise to 
a dramatic deviation of a partial sum $B_N^{(a)}(\theta)$
from the numerically calculated values of the baryon number already at rather small values of $\theta_R$, see Fig.~\ref{fig:overall} below.

Fortunately, the series~(\ref{eq:Fourier_expansion}) 
can be summed up analytically in both CEM and RFM cases,
which can be used to perform the analytic continuation to the entire complex plane of $\theta$.
The summation was first presented in~\cite{Vovchenko:2018zgt} 
for the CEM, here we rewrite it as an analytic 
function of the complex variable $\theta$ using 
the parameters $b$ and $q$ for QCD with $N_f$ quark flavors and $N_c$ colors:
\beq
B_{CEM}(\theta) = {b\over 2q}\left\{ \ln {1+q\exp(\theta N_c) \over 1+q\exp(\,-\;\theta N_c)}
  + {6\over \pi^2 (N_c^2-1)}
\left[\mathrm{Li}_3\Big(-q e^{-\theta N_c}\Big) - \mathrm{Li}_3\Big(-q e^{\theta N_c}\Big) \right] 
  \right\} \ .
\label{CEM_summed}
\eeq
In the limit of free massless $N_f$ quark flavors 
one has $q=1$ and
\beq
b={2\over 3}\; {N_f (N_c^2-1)\over N_c^3}\; {N_s^3\over N_t^3}\;,
\eeq
and Eq.~\eqref{CEM_summed} reduces to the baryon number 
of the free quark gas.

In Fig.~\ref{fig:CEM_vs_lattice2} we plot 
the baryon number $B(\theta)$ divided by $\theta$ versus 
$\theta^2$ including both imaginary ($\theta^2 < 0 $) 
and real ($\theta^2 > 0 $) values of $\theta$. 
In the left panel our lattice results for $T=227$~MeV 
and $T=265$~MeV are depicted together with  fits to the CEM, 
Eq.~(\ref{CEM_summed}). The error band in this figure 
is determined from the condition that the function 
(\ref{CEM_summed}) at each particular value of $\theta$ 
is considered as the function of two correlated random 
variables $b$ and $q$; in so doing, we employ bootstrapping. 
One can see a very good agreement between 
the CEM prediction and our 
lattice data for real values of $\theta$ 
over the full range of the explored chemical potentials covering $0 < \mu_q/T < 3.5$ or $ 0  < \mu_q < 800$~MeV 
for $T=227$~MeV and $0 < \mu_q/T < 2.4$ or 
$ 0  < \mu_q < 640$~MeV for $T=265$~MeV. This agreement 
is especially good for $T=265$~MeV. It is important 
to extend our analysis to other values of $T$ in order 
to understand whether the observed agreement between 
the CEM and lattice data is common to all $T<T_{RW}$. 

We shall emphasize that the function $B_{CEM}(\theta)$ 
while providing good extrapolation to real values of $\theta$ over a wide range has a drawback: it has branch-cut singularities in $\theta$ complex plane along the lines 
$$
\big\{ \theta: \theta_I=(2k+1)\pi,\ k\in \mathbb{Z},\ |\theta_R|>-\ln q\;\big\} \;,
$$ 
that is, it is not meromorphic.  Therefore, the respective 
grand canonical partition function $Z_{GC}^{CEM}(\theta)$ 
is not an entire function, it has branch-cut 
singularities~\cite{Taradiy:2019taz} 
along the negative real semiaxis in the $\xi$ complex plane: 
$\ds -\;{1\over q}<\mathrm{Re}\xi<-q$.

In the left panel we also show results of the fit 
to Eq.~(\ref{eq:ff_for_Z}) and the respective analytic 
continuation. One can see that at $T=265$~MeV this approach 
to analytic continuation works quite well providing 
reasonable analytic continuation up to $\theta \sim 0.5$. 
At  $T=227$~MeV result of analytic continuation is not so good. 
We believe that this is due to insufficient statistics 
we have at imaginary $\theta$
for this temperature. We will provide additional arguments in favor of this approach to analytic continuation in the next subsection. 

We also perform the summation of the Fourier series in the RFM. This summation, which has not been presented in the prior literature, has the form 
\bea\label{eq:RFM_summed_series}
\hspace*{-3mm}-\;\imath B_{RFM}(\theta)\Big|_{\theta_R=0}\!\!\hspace*{-3mm} &=& \!\!\!d\; \left\{
\left( {\ds \pi^2(N_c^2-1)\over 6} +\kappa^2\right) 
\left[ {\theta_I N_c\over 2} - \left(\beta\left({1\over \kappa}\right) -{\kappa\over 2}\right)\sin \left( {\theta_I N_c\over \kappa}  \right)\right.\right.  \\ \nonumber
&&\left. - \left. {1\over 2} \int_0^{\theta_I N_c} dt\; \tan{t\over 2} \sin{\theta_I N_c-t\over \kappa}
\right] +{\pi^2\over 12} \left(\theta_I N_c - {(\theta_I N_c)^3\over \pi^2} \right) - \kappa 
\int_0^{\theta_I N_c} \ln \left(2\cos{t\over 2}\right) dt \right\}
\nonumber 
\eea
where $\ds \beta(z)={1\over 2} \left(\psi\left({z+1\over 2}\right)-\psi\left({z\over 2}\right)\right)$ and  $\ds \psi(z)={1\over \Gamma(z)}\,{d\Gamma(z)\over dz}$ is the logarithmic derivative of the 
Gamma function  $\Gamma(z)$. The limit of free massless quarks with $N_c$ colors is approached when $\kappa=0$ and $\ds d={4N_f \over \pi^2 N_c^3}\;{N_s^3\over N_t^3}$.

The above expression represents the imaginary part of the function
\bea\label{eq:RFM_summed_series_2} B_{RFM}(\theta) &=& 
d\; \left\{ \left( {\ds \pi^2(N_c^2-1)\over 6} +\kappa^2\right) 
\left[ {\theta N_c\over 2} - \left(\beta\left({1\over \kappa}\right) -{\kappa\over 2}\right)\sinh \left( {\theta N_c\over \kappa}  \right) \right. \right.
    \\ \nonumber
&& + \left. \left. \,{1\over 2} \int_0^{\theta N_c} dt\; \tanh{t\over 2} \sinh{\theta N_c-t\over \kappa}
\right] + {\pi^2\over 12} \left(\theta N_c + {(\theta N_c)^3\over \pi^2} \right) - \kappa 
\int_0^{\theta N_c} \ln \left(2\cosh{t\over 2}\right) dt \right\}
\nonumber 
\eea
at the imaginary values of $\theta$. 

As clearly seen in the right panel of Fig.~\ref{fig:CEM_vs_lattice2}, in contrast to the CEM, 
the RFM significantly deviates from the lattice data 
at real $\theta$. 
Qualitatively, $B_{RFM}(\theta_R)$ does capture some of the features in the lattice data at $T=265$~MeV. However, it exhibits a substantially different qualitative behavior at $T=227$~MeV where $B_{RFM}(\theta_R)$ 
is a strongly convex function, whereas the data show a slight concavity. Let us note that the function (\ref{eq:RFM_summed_series}) 
involves nonanalyticity of the type $\ds \left({\pi\over N_c}-\theta_I \right) \ln \left({\pi\over N_c}-\theta_I \right)$: its first derivative tends to infinity as $\theta_I\to \pi/N_c$. 
This observation is of relevance, as it is contrary to the expectation that the quark density at $T<T_{RW}$ is
an analytic function on some domain of the $\theta$ plane containing the imaginary axis.
This may explain the significant deviations of the RFM from the lattice data at $T < T_{\rm RW}$.

To conclude, the analytic continuation
of the quark density to real quark chemical potentials 
provided by the CEM shows a very good agreement 
with the lattice data contrary to the case of RFM.
This observation is similar to the conclusions 
of Ref.~\cite{Vovchenko:2019vsm} regarding 
the description of baryon number susceptibilities 
in (2+1)-flavor QCD, where the CEM was also found 
to provide a more accurate description of the lattice data.

\subsection{$Z_n$ in the CEM and RFM and canonical approach to the extrapolation } 
\label{sec:find_Z}

In this subsection, we first discuss the computation 
of $Z_n$ for the CEM and RFM and their properties 
and then discuss the analytic continuation 
via Eq.~(\ref{eq:ff_for_Z}) using these $Z_n$. 
As explained in Appendix~\ref{sec_App1}, 
$Z_n$ can be easily evaluated when the Fourier 
coefficients $a_n$ are known from 
the fit by Eq.~(\ref{eq:Fourier_expansion}) 
or from the fit based on a model 
(see Tables~\ref{tab:compare_Z_n_a_T14} 
and~\ref{tab:compare_Z_n_a}).  

%%%%%%%%%%%% BEGIN FIGURE 2 %%%%%%%%%%%%%%%%%
\begin{figure}[htb]
\begin{center}
\includegraphics[width=15.0cm]{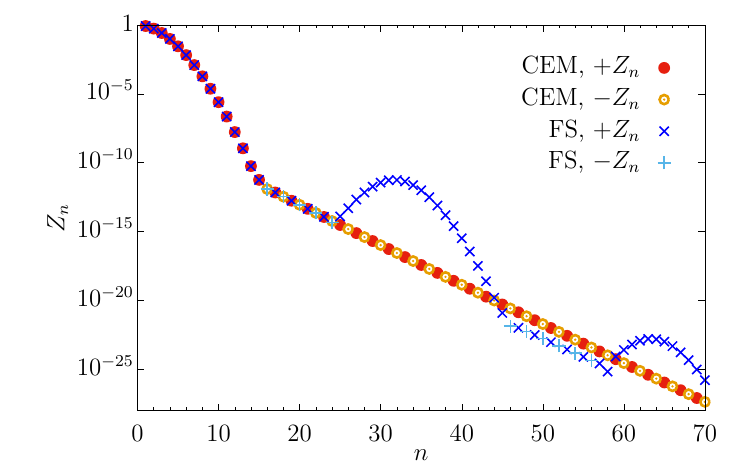}
\caption{ Behavior of $Z_n^{\mathrm{CEM}}$ at $T=265$~MeV.
Shown are the cases when the baryon number is computed by Eq.~(\ref{CEM_summed}) (CEM) and by the truncated  Fourier series Eq.~(\ref{eq:B_CEM_appr}) with $N=31$ (FS). }
%series of 31 sines (SS).  }
\label{fig:Zn_vs_n}
\end{center}
\end{figure}
%%%%% END FIGURE 2 %%%%%%%%%%%%%%%%%%%%%%%%%%%%%%%%%%%

Alternatively, the coefficients $Z_n$ can be computed  
from a given expression for the baryon density using 
the integration method discussed in Ref.~\cite{Bornyakov:2016wld}. 
Positivity of $Z_n$ provides a test of whether 
the function $\tilde{B}(\theta_I)$ corresponds to a physical system.
Also, in case where the analytic continuation of $\tilde{B}(\theta_I)$ is not evident, $Z_n$ can be used to determine the functions 
\beq\label{eq:rho_canon} 
B^{(Z)}_N(\theta) = \frac{2\sum_{n=1}^N n Z_n \sinh(2n\theta)}{1+2\sum_{n=1}^N Z_n \cosh(2n\theta)} %\;,
\eeq
providing an approximation to $B(\theta)$ at real $\mu_q$. 
Here we want to use $Z_n$ computed for the CEM 
to demonstrate further potential of this approach 
which we call the canonical approach.

In a particular case when the baryon number is given by Eq.~(\ref{eq:Fourier_expansion})
with known Fourier coefficients $a_n$ 
\beq
\frac{Z_{GC}(\theta_I)}{Z_{GC}(0)} = \exp \left( -N_c\int_0^{\theta_I} B^{(a)}_N(x)\; dx \right)
=\exp \left( \sum_{n=1}^N {a_n\over n}\;\Big(\cos(nN_c\theta_I) - 1 \Big) \right)\;
\label{eq:ZGCintegr}
\eeq
and one can use Eq.~(\ref{Fourier}) and  relation
$Z_n=Z_C(2n) / Z_C(0)$  to evaluate $Z_n$. 
The comparison between the results obtained by 
the integration method and by the method 
of Appendix \ref{sec_App1} also provides 
a cross-check of the accuracy of the computations 
of $Z_n$. We find that the respective results 
coincide with a precision of up to $20$ significant digits.

Fig.~\ref{fig:Zn_vs_n} shows the behavior of $Z_n$ in the CEM 
(denoted below as $ Z_n^{\mathrm{CEM}}$)  for $T = 265$~MeV.  
$Z_n^{\mathrm{CEM}}$ were computed by 
the integration method using the analytical expression~(\ref{CEM_summed}) which represents the sum of the entire infinite Fourier series with the CEM coefficients~\eqref{eq:CEM_coeffs}. For comparison, we show results obtained with the baryon number approximated by the truncated Fourier series
\beq
\tilde{B}_N^{(a)CEM}(\theta_I) = \ds \sum_{n=1}^N \acem \sin\left({2 n \theta_I}\right) 
\label{eq:B_CEM_appr}
\eeq
with $N=31$  and $\acem$ given by~\eqref{eq:CEM_coeffs}.
We found that $Z_n^{\mathrm{CEM}}$ computed with exact 
and approximate $\tilde{B}_{CEM}(\theta_I)$ 
agree with each other for sufficiently small $n$, $n \leq 24$.
Differences occur at larger $n$, which are attributed to the artefact of using a truncated Fourier series. When we take $N=101$ in Eq.~(\ref{eq:B_CEM_appr}), we find agreement for the full range of $n$ shown in Fig.~\ref{fig:Zn_vs_n}. 

From the physics point of view,  $Z_n^{\mathrm{CEM}}$ 
exhibit a regular behavior at $n \leq 15$, 
where they are all positive and demonstrate fast decreasing.
However, at $n \geq 16$ negative values 
of $Z_n^{\mathrm{CEM}}$ are obtained which is unphysical. 
Moreover, they decrease as $e^{-\alpha n}$ ($\alpha=-\ln q$), which 
implies that the series in Eq.~(\ref{eq:second_limit}) (see below) 
has a finite radius of convergence
  associated with the 
branch cut singularities~\cite{Taradiy:2019taz} of the CEM partition function along the negative real semiaxis at $\ds \xi \leq { -1\over q }$ and $\ds {-q \leq \xi \leq 0}$. 

For $T = 227$~MeV a qualitatively similar behavior of $Z_n^{\mathrm{CEM}}$ is observed. In particular, the first negative $Z_n^{\mathrm{CEM}}$ appears at $n=12$.

In Tables~\ref{tab:compare_Z_n_a_T14} 
and~\ref{tab:compare_Z_n_a}  we compare  $Z_n$ computed via direct fit, Eq.~(\ref{eq:ff_for_Z}), with $Z_n^{\mathrm{CEM}}$. One can see very good agreement between these sets of $Z_n$, especially for $T=265$~MeV.

To solve the problem of negative $Z_n^{\mathrm{CEM}}$, 
the expressions  (\ref{eq:CEM_coeffs}) 
for the coefficients $a_n^{\mathrm{CEM}}$ should be modified. 
This modified version of the CEM should take into account
attractive interaction and the finite-volume effects. 
Yet another possibility to avoid negative $Z_n^{\mathrm{CEM}}$ 
is to follow Refs.\cite{Roberge:1986mm,Bornyakov:2022blw} 
and use saddle-point approximation of the Fourier integral 
(\ref{Fourier}) instead of its precise numerical evaluation. 
However, a proper justification and physical interpretation 
of such approach should be the subject of another study.

%%%%%%%%%%      BEGIN FIGURE 3        %%%%%%%%%%%%%%%%%%%%%
\begin{figure}[tbh]
\vspace*{-18mm}
\hspace*{-9mm}
\includegraphics[width=9.2cm] {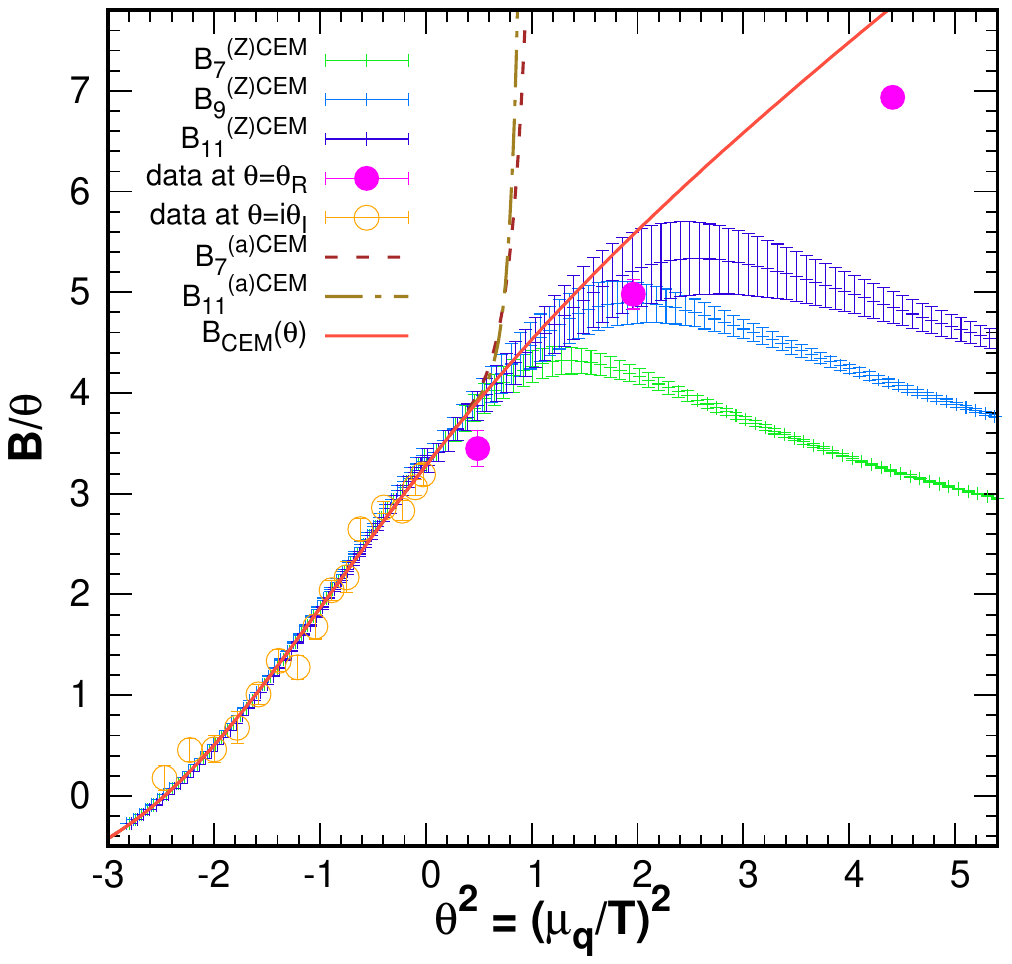}\hspace*{-10mm}
\includegraphics[width=9.2cm]{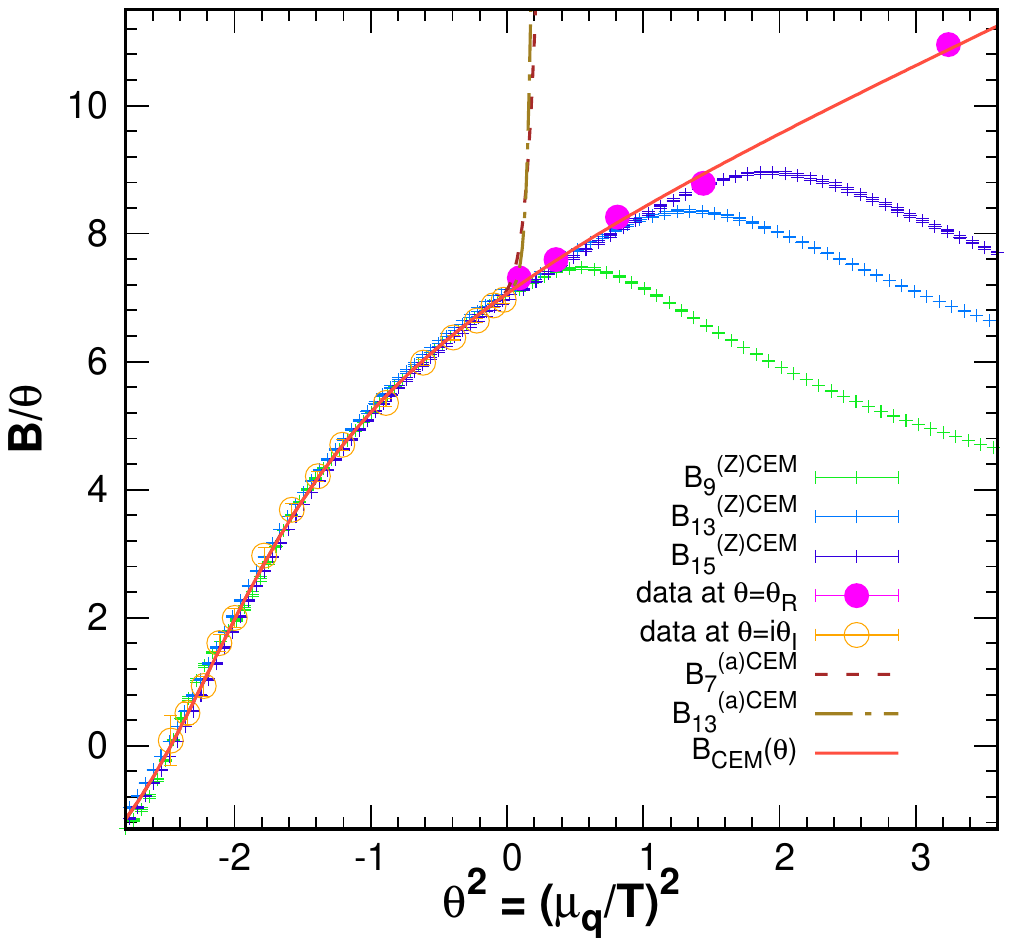}
\vspace*{-20mm}
\caption{The functions $B^{(a)CEM}_N(\theta)/\theta$ and 
$B^{(Z)CEM}_N(\theta)/\theta$
obtained in the CEM at various $N$ are plotted versus $\theta^2$ at $T=227$~MeV (left panel) and $T=265$~MeV (right panel) together with the respective lattice data and CEM baryon number Eq.~(\ref{CEM_summed}). Error bands for $B^{(a)CEM}_N(\theta)/\theta$ and $B_{CEM}(\theta)$ are not plotted for clarity, error bands for $B_N^{(Z)CEM}(\theta)/\theta$ in the right panel can be seen on screen by zooming in. 
}
\label{fig:overall}
\end{figure}
%%%%%%%%%%      END FIGURE 3        %%%%%%%%%%%%%%%%%%%%%

The expectation value of the baryon number at physical values 
of the quark chemical potential can be obtained not only  
via Eq.~(\ref{CEM_summed}), but also through the use of $Z_n$ via the formula
\beq
B^{(Z)}(\theta_R) = \lim_{N\to\infty} B^{(Z)}_N(\theta_R) \equiv \lim_{N\to\infty} {2 \sum_{n=1}^N n Z_n \sinh(2n\theta_R) \over 1 + 2 \sum_{n=1}^N  Z_n \cosh(2n\theta_R)} \;.
\label{eq:second_limit}
\eeq

In Fig.~\ref{fig:overall} we compare the functions
\begin{displaymath}
\frac{1}{\theta}B^{(a)CEM}_N(\theta)=\frac{1}{\theta}\sum_{n=1}^N \acem \sinh(\theta) = \left\{  \begin{array}{ll}
\ds \frac{1}{\theta_R}\sum_{n=1}^N \acem \sinh(\theta_R) \quad & \mbox{if}\quad \theta_I=0 \\[4mm]
\ds \frac{1}{\theta_I}\sum_{n=1}^N \acem \sin(\theta_I) \quad & \mbox{if}\quad \theta_R=0 \\
\end{array}
\right.
\end{displaymath}
and 
\begin{displaymath}
\frac{1}{\theta}B^{(Z)CEM}_N(\theta)=\frac{1}{\theta}\frac{2\sum_{n=1}^N n Z_n^{CEM} \sinh(2n\theta)}{1+2\sum_{n=1}^N Z_n^{CEM} \cosh(2n\theta)}
= \left\{  \begin{array}{ll}
\ds \frac{1}{\theta_R}\frac{2\sum_{n=1}^N n Z_n^{CEM} \sinh(2n\theta_R)}{1+2\sum_{n=1}^N Z_n^{CEM} \cosh(2n\theta_R)} \quad & \mbox{if}\quad \theta_I=0 \\[7mm]
\ds \frac{1}{\theta_I} \frac{2\sum_{n=1}^N n Z_n^{CEM} \sin(2n\theta_I)}{1+2\sum_{n=1}^N Z_n^{CEM} \cos(2n\theta_I)} \quad & \mbox{if}\quad \theta_R=0 \\
\end{array}
\right.
\end{displaymath}
obtained in the CEM at various $N$. At both temperatures
we see that both $B^{(a)CEM}_7(\theta)$ and $B^{(a)CEM}_{13}(\theta)$
depart from the lattice data dramatically at real $\theta$
starting from a small value $\theta_R$ ($ \theta_R^2 \sim 0.6$ for $T=227$~MeV and  $\theta_R^2 \sim 0.12$ for $T=265$~MeV). The point is that the series
\beq 
\sum_{n=1}^\infty \acem \sinh(2n\theta)
\label{eq:CEM_sinhs_series}
\eeq 
diverges at  $\ds \theta_R > -\;{\ln(q)\over 2}\approx 0.81$ 
for $T=227$~MeV and $\theta_R>0.33$ for $T=265$~MeV,
as it follows from formulas (\ref{eq:CEM_coeffs}) and
(\ref{eq:diverg_of_sinhs}). 

As one can see from Fig.~\ref{fig:overall}
the functions  $B^{(Z)CEM}_N(\theta)$ approximate well  
the lattice data over a broad  range of   $\theta_R$
and this range increases with $N$. However, 
there exists a maximum value of $N=N_{max}$ 
such that $Z_n$ at $n> N_{max}$ are not properly 
extracted from the data, which is indicated 
by alternating sign and a slow decrease 
of the absolute value of 
$Z_n$ at $n>N_{max}$ as is shown in Fig.~\ref{fig:Zn_vs_n}.
The values of $N_{max}$ and the corresponding ranges of 
$\theta_R = \mu_q/T$ are as follows:
\begin{itemize}
\item $|\theta| \lesssim 1.3 \quad (|\mu_q| \lesssim 295$~MeV) \ at $N=N_{max}=11$ for $T=227$~MeV, 
\item $|\theta| \lesssim 1. 2 \quad (|\mu_q| \lesssim 320$~MeV) \ at $N=N_{max}=15$ for $T=265$~MeV. 
\end{itemize}
Since  $N_{max}$ values specified here give the upper
bounds on the ranges of values of $n$ where all $Z_n$ computed for the CEM are positive, we consider
the respective values of $\mu_q/T$
as the upper bounds on the domain of $\mu_q/T$ where
the corrections to the CEM can be neglected.

We leave beyond the scope 
of the present study a question on the 
behavior of $B_N^{(Z)CEM}(\theta)$ at $N>N_{max}$
because they do not correspond to a physical system. The relation between the appearance of the negative $Z^{CEM}_n$ and the properties of $B_{CEM}(\theta)$ should
be studied in the future.

The analytic continuation of the Fourier series in RFM
to real values of $\theta$ is more subtle.
This series
\beq
\sum_{n=1}^\infty \arfm \sinh(2n\theta)
\eeq 
(analogous to the series~(\ref{eq:CEM_sinhs_series}) in CEM)
diverges at all real $\theta\!\neq\! 0$ because of only power-like decrease of $\arfm$ with $n$.

%%%%%%%%%%%%%%% BEGIN FIGURE  4 %%%%%%%%%%%%%%%%%%%%%%%
\begin{figure}[hhh]
\vspace*{-15mm}
\begin{center}
\hspace*{-8mm}\includegraphics[width=8.9cm]{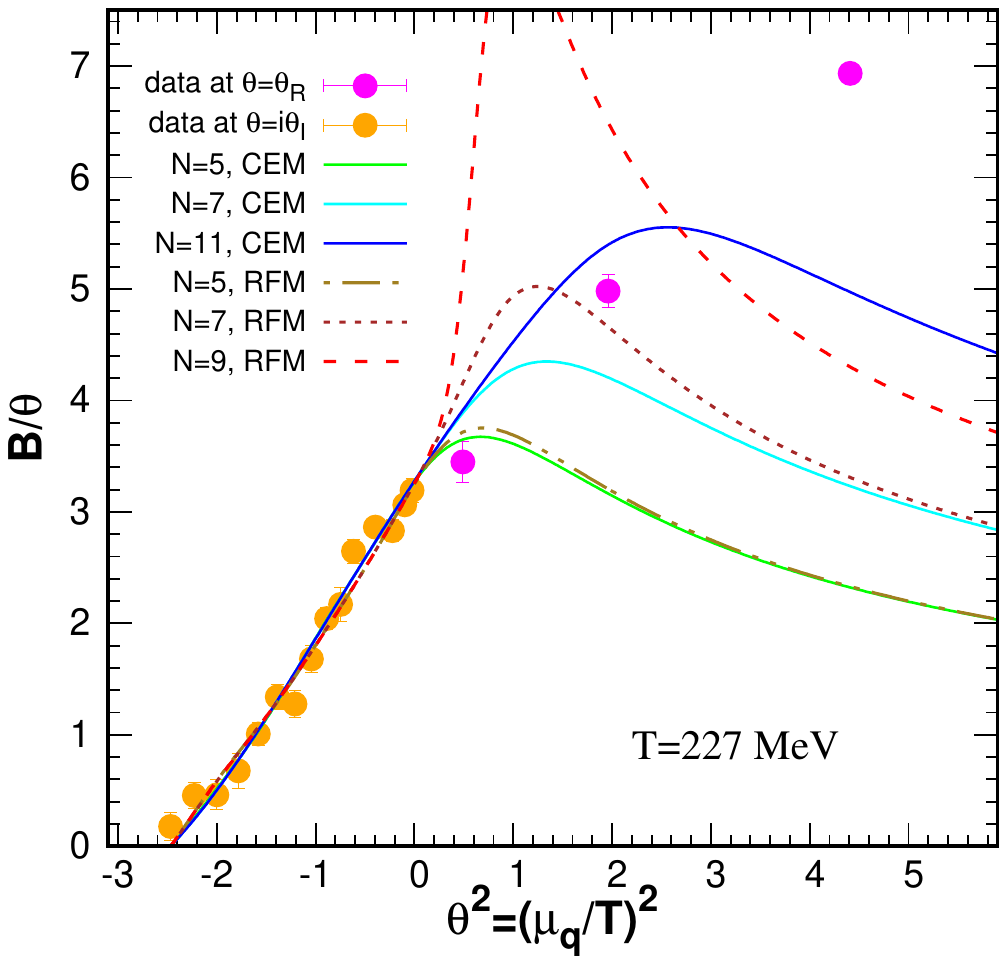}
\hspace*{-12mm}\includegraphics[width=8.9cm]{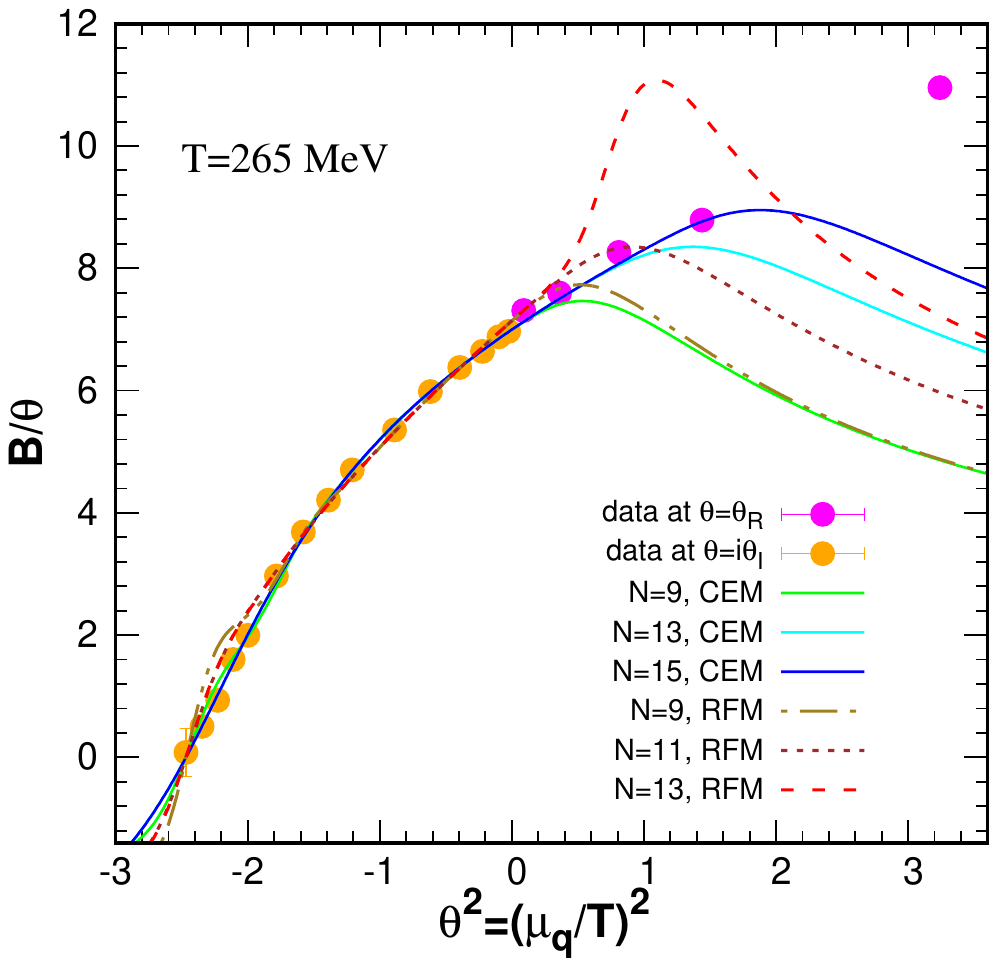}
\vspace*{-20mm}
\caption{ The functions $B^{(Z),CEM}_N$ obtained in the CEM (solid lines) are compared with those obtained in the RFM (dashed lines) at various $N$: $T=227$~MeV (left panel) and $T=265$~MeV (right panel).}
\label{fig:compare}
\end{center}
\end{figure}
%%%%%%%%%%%%%%% END FIGURE  4 %%%%%%%%%%%%%%%%%%%%%%%

In Fig.~\ref{fig:compare} we compare the quantities $B^{(Z)CEM}_N$ obtained in the CEM 
with those obtained in the RFM at real values of $\mu_q$ ($\theta^2\geq 0$). 
We are interested in the range of real $\mu_q$, 
where the plots of $B^{(Z)}_N$ come close 
to the lattice data. This range extends with an increase 
of $N$ in the case of CEM and vanishes at all $N$ 
in the case of RFM. 

\section{Analytic continuation of the quark number density: high temperatures}
\label{sec:highT}

At high temperatures ($T>T_{RW}$) the conventional method is 
to use polynomial fit functions, namely,
one performs a fit by a polynomial of degree $2N_p-1$ to data for the baryon number
at imaginary values of $\theta=i\theta_I$ and then makes the analytic 
continuation to real values of $\theta$~(see e.g. \cite{Takahashi:2014rta,Bornyakov:2016wld}). 
In this case, $N_p$ determines the number of fit parameters.
This number is less than the order of the polynomial $2N_p-1$ because all even powers vanish due to the CP-symmetry.
We analyze two values of temperatures in this regime: $T = 398$~MeV and $530$~MeV.

\subsection{Extrapolation based on power series expansion}

%--------------------------------------------------------------------%%%%% FIGURE
\begin{figure}[tbh]
\vspace*{-25mm}
\hspace*{-7mm}\includegraphics[width=9.8cm]{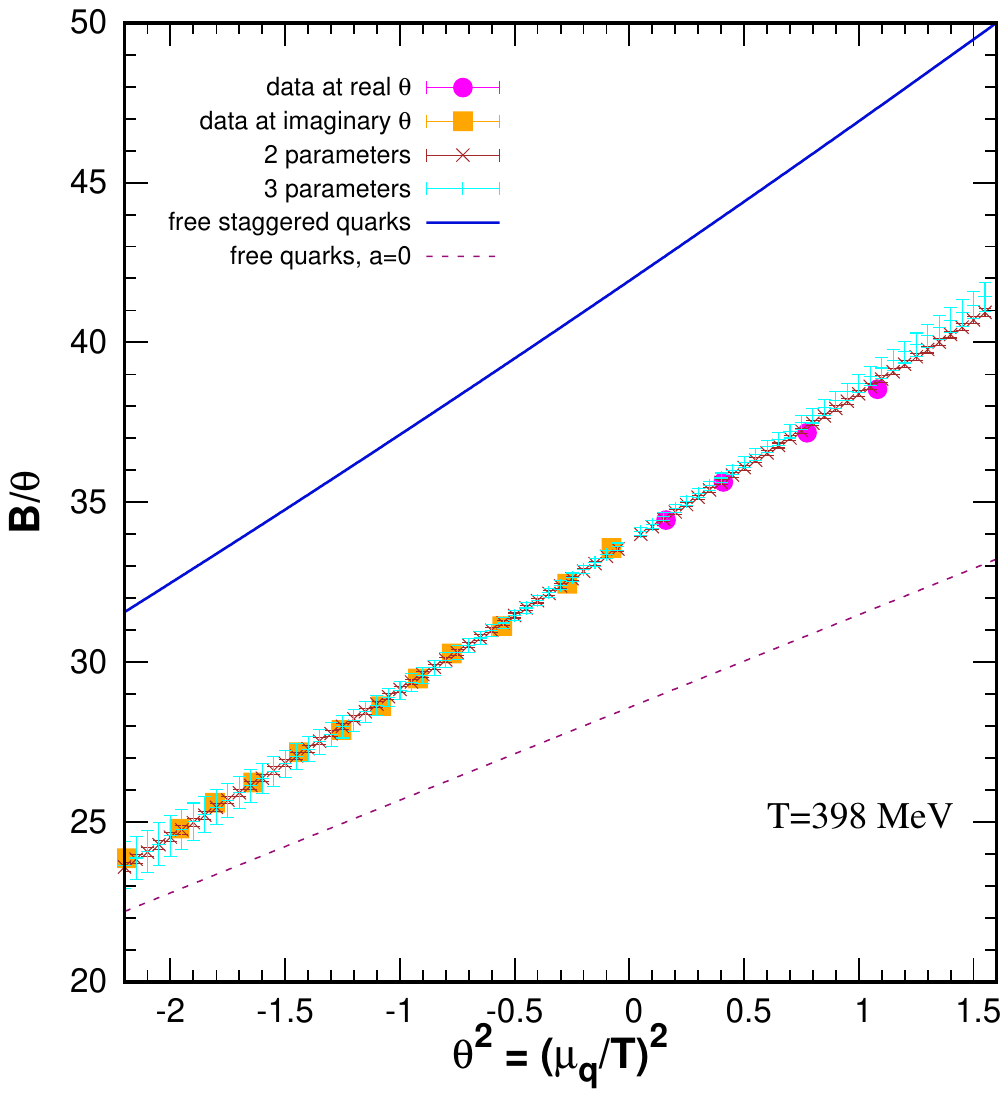}\hspace*{-9mm}
\includegraphics[width=9.8cm]{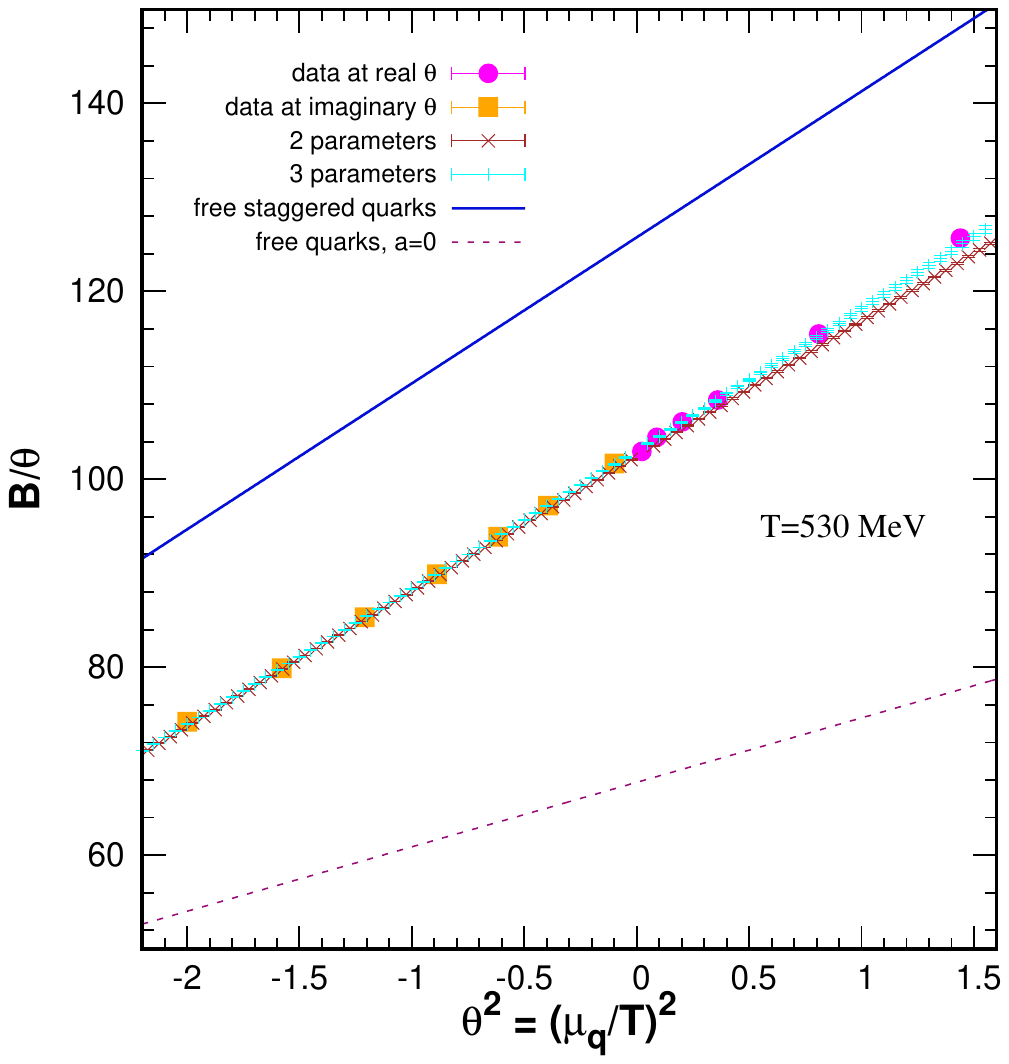}
\vspace*{-28mm}
\caption{Analytic continuation of the baryon number 
from imaginary to real $\theta$ with the use 
of the power fit (\ref{eq:pow_fit}) at $T=398$~MeV 
(left panel) and $T=530$~MeV (right panel). 
The baryon number for free quarks (dashed line) and
for free lattice  staggered
quarks (solid line) are also shown.
The error bands are obtained by the bootstrap method.
}
\label{fig:extrap_pow_fit}
\end{figure}
%%%%% END FIGURE  
%--------------------------------------------------------------------

Thus, we fit the baryon number at imaginary chemical potential by the function
\beq\label{eq:pow_fit}
\tilde B^{(c)}(\theta_I) = \sum_{n=1}^{N_{p}} (-1)^{n+1} c_n   \;\theta_I^{2n-1}
\eeq
using the values $N_{p}=2$ and 3. 
The fit parameters are $c_n$ and their resulting values are shown in Table~\ref{tab:pow_fit}.
Fig.~\ref{fig:extrap_pow_fit} depicts the analytic continuation to the domain of real $\theta$ using the values of the extracted fit parameters.
The error bands in Fig.~\ref{fig:extrap_pow_fit} are obtained by the bootstrap method. 
Comparison with the lattice data at real values 
of $\theta$ indicates that the most accurate analytic 
continuation is achieved by using 
a polynomial of the fifth degree at $T=530$~MeV 
and a polynomial of the third degree at $T=398$~MeV.

%%%%%%%%%%%%%%%%%%%%%%%% Table IV %%%%%%%%%%%%%%%%%%%%%%%%%%%%%%
\begin{table}[hhh]
\bc
\begin{tabular}{|c|c|c|c|c|c|} \hline
~$T\;$(MeV)~ & ~$N_{param}$~   & $\chi^2/N_{dof}$ & ~$c_1$~  &  ~$c_2$~ & ~$c_3$~ \\ \hline
 398 &  2  &  0.83  &  33.698(92)  & 4.548(76) &  -          \\
 398 &  3  &  0.64  &  33.83(16)   & 4.88(33)  &  0.15(14) \\
% 398 &  4  &  0.70  &  33.86(22)   & 5.07(92)  &  0.4(1.0)  \\ \hline
  398 &  -   & free quarks &  28.58  & 2.90  &      \\ 
 398 &  3   & free stag. quarks &  41.94  & 4.812  &  0.094     \\   \hline
 530 &  2  &  4.19  &  102.859(72) & 14.467(66) &  -          \\
 530 &  3  &  0.83  &  103.18(13) & 15.29(27)  &  0.38(12)  \\
% 530 &  4  &  0.97  &  103.12(18)  & 14.98(68)  &  0.03(71)   \\
 \hline 
 530 &  -   &  free  quarks & 67.75   & 6.86  &      \\    
 530 &  3   & free stag. quarks & 125.83   & 15.45  &  0.297    \\
\hline
\end{tabular}
\ec

\caption{The results of the fit by Eq.~(\ref{eq:pow_fit}).
%$c_4$ is poorly determined and is not shown. 
The results for the free quarks and for the free lattice staggered quarks are given for comparison.}
\label{tab:pow_fit}
\end{table}
%%%%%%%%%%%%%%%%%%%%%% end Table IV   %%%%%%%%%%%%%%%%%%%%%%%%%%%%%%%%%%
To check the role of quark interactions, 
we compare our results with the respective results for the free massless quarks for the same volume and temperature.
In the continuum limit, the free-quark partition function 
has the form
\beq
Z_{GC}^{(free)}(\theta) = A\exp \left[ c\left( \theta^2 + {\theta^4\over 2\pi^2}\right) \right]\; 
\eeq
where $\ \ds c={g_f VT^3\over 12}\ $ (in the case of lattice QC$_2$D with two flavors, $\ds c={2 N_s^3\over 3 N_t^3}\;$), and $\ds A=\exp \left({7c\pi^2\over 30}\right)\;$ . 
The baryon number is then determined as
\beq\label{eq:baryon_number_free_quarks}
B^{free}(\theta) =  c \left(\theta + {\theta^3\over \pi^2}\right)~.
\eeq
This corresponds to a polynomial of degree 3 ($N_{p}=2$) in Eq.~(\ref{eq:pow_fit})
with coefficients $c_1^{free}$ and $c_2^{free}$.
The numerical values of these coefficients 
are shown in Table~\ref{tab:pow_fit} for comparison 
with lattice results.
Furthermore, we compute the baryon number 
for the lattice staggered-fermion action used in our study,
the respective coefficients $c_{n,stag}^{free}$ 
are presented in  Table~\ref{tab:pow_fit}.  
We present $B^{free}(\theta)$ and $B^{free}_{stag}(\theta)$ in Fig.~\ref{fig:extrap_pow_fit}.
Both from the Table~\ref{tab:pow_fit} and from 
the Fig.~\ref{fig:extrap_pow_fit} one can see that, 
owing to discretization effects, $B^{free}_{stag}(\theta)$ 
is quite different from $B^{free}(\theta)$ for our lattices with $N_t=6$ and 8.  
For this reason, we compare our results with 
$B^{free}_{stag}(\theta)$ rather than with $B^{free}(\theta)$.

It follows from Table~\ref{tab:pow_fit} that the coefficient $c_2$ agrees with 
$c_{2,stag}^{free}$ within error bars (this is also true for $c_3$ but its error is large), while $c_1$ is still 20\% lower than $c_{1,stag}^{free}$. This can be also seen  from Fig.~\ref{fig:extrap_pow_fit}. 
 The large value of $c_{3,stag}^{free}$ 
for $T=530$~MeV indicates importance 
of the term $\theta^5$ for this temperature.
However, this term is needed at $T=530$~MeV 
and should be omitted at $T=398$~MeV not for any physical 
reason but simply because the discretization effects 
for $N_t=6$ are greater than for $N_t=8$. 

Thus, the baryon number in the free quark limit  $B^{free}_{stag}(\theta)$ 
is higher than that for interacting
quarks; that is, the presence of interactions decreases the baryon number 
at given parameters  $V,T,\mu_q$.
This is in a qualitative agreement with (2+1)-flavor QCD with physical quark masses, where lattice QCD calculations indicate that the baryon number is suppressed compared to the free quark limit at similar temperatures and chemical potentials~\cite{Bazavov:2017dus}.  We conclude that the fit (\ref{eq:pow_fit})  provides a good analytic continuation 
at the considered temperatures but we need 
 a finer lattice  for better understanding of the quark interaction effects.
%%%%%%%%%%%%%%%%%%%%%%%%%%

\subsection{Other schemes}

Despite success of the analytic continuation based on Eq.~(\ref{eq:pow_fit}) it is worth to explore other possibilities for performing the analytic continuation at $T > T_{\rm RW}$. We expect that when temperatures get closer to $T_{RW}$ than we study in this paper, the functional dependence $B(\theta)$ becomes more complicated due to increased interaction effects.

We begin with a consideration of the trigonometric Fourier series 
expansion of the baryon number over the segment 
$\ds -\;{\pi\over 2} <\theta_I \leq {\pi\over 2}$,
focusing on $T = 398$~MeV.
We utilize the truncated Fourier series
\beq\label{eq:sin_fit}
\tilde{B}^{(a)}_{N_p}(\theta_I) = \sum_{n=1}^{N_{p}} a_n  \sin\big(2n\theta_I\big)~.
\eeq
The fit quality is unsatisfactory,
giving $\ds {\chi^2\over N_{d.o.f.}}>200$ for $N_p$ as large as $N_{p}=8$. 
The reason is rather obvious. The quark number 
density is a discontinuous function of
$\theta_I$ at the edges of the interval $\ds -\;{\pi\over 2} <\theta_I \leq {\pi\over 2}$
and the Fourier series converges very slowly 
to a discontinuous function.

A different possibility is to use the canonical formalism and the associated formula (\ref{eq:ff_for_Z}), which has been shown to work reasonably well at $T < T_{\rm RW}$.
However, a fit based on this formula also does not give
a reasonable result,  we obtain $\ds {\chi^2\over N_{d.o.f.}} > 1000$ and the results of the fit are heavily dependent on initial values of the parameters. 
Thus, we shall consider other possibilities.

\begin{figure}[tbh]
\vspace*{-20mm}
\begin{center}
\hspace*{-12mm}\includegraphics[width=9.3cm]{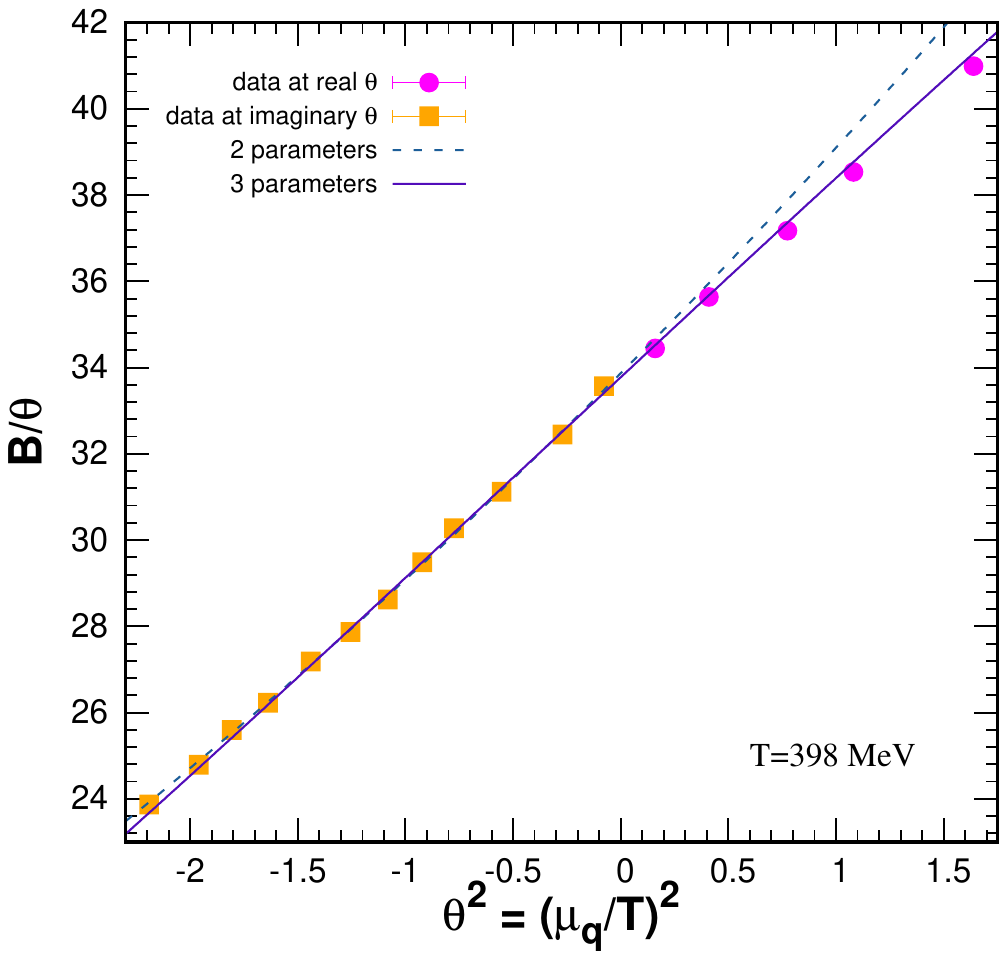} \hspace*{-9mm}
\includegraphics[width=9.3cm]{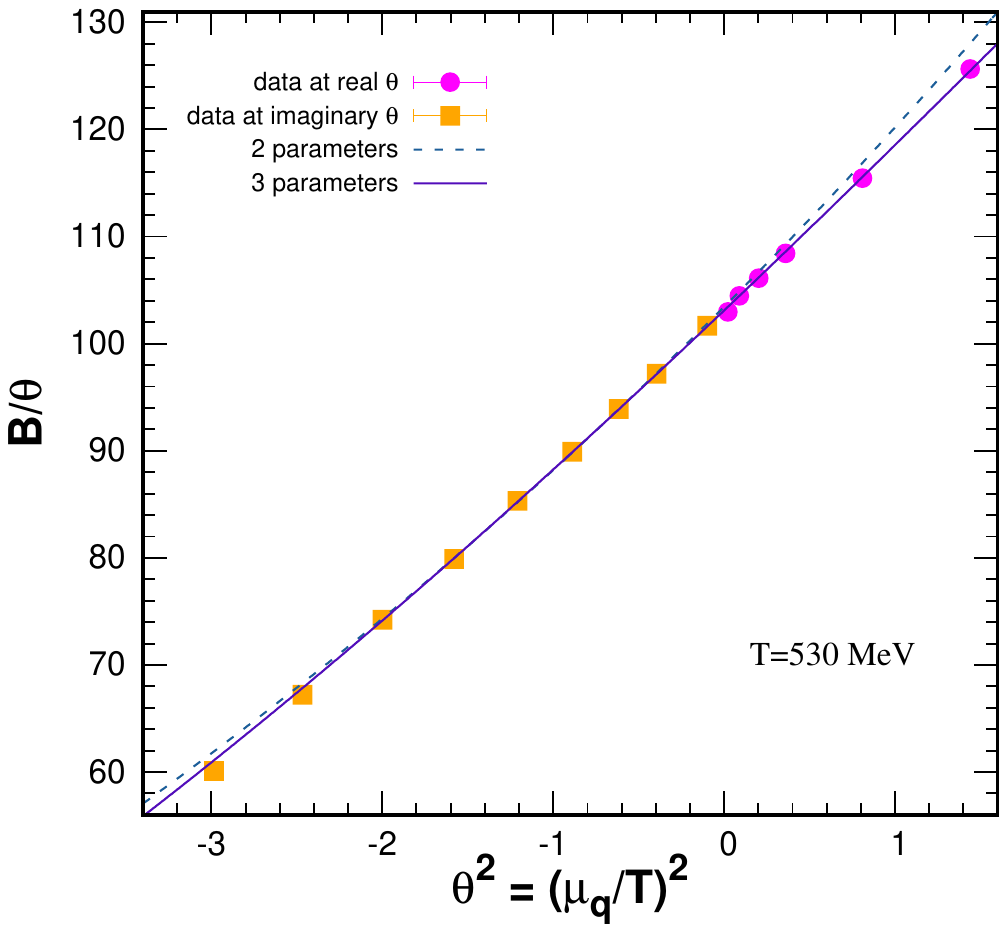}
\vspace*{-24mm}
\caption{ Two different procedures of analytic continuation are compared: the two-parameter fit is based on formula (\ref{eq:sin_fit_irrat}),
the three-parameter fit is based on formula (\ref{eq:canon_08}) associated with the canonical formalism. }
\label{fig:qdr_Nt08_sin_tot_02}
\end{center}
\end{figure}

We observe that the baryon number can be fitted quite well
using the function
\beq\label{eq:sin_fit_irrat}
\tilde{B}(\theta_I) \simeq  c \sin\big(w\theta_I\big)\;,
\eeq
where $c$ and $w$ are fitting parameters. 
The fit results are presented in Table~\ref{tab:3fits}.
Eq.~(\ref{eq:sin_fit_irrat}) can be used to perform the analytic continuation to the real values of $\theta$, the result is shown in Fig.~\ref{fig:qdr_Nt08_sin_tot_02}.
%----------------------------------------%%%%% FIGURE
%%%%% END FIGURE
%---------------------------------------------------------
One can see that the analytic continuation based on Eq.~\eqref{eq:sin_fit_irrat} yields a reasonably good agreement with the lattice data, but systematically overestimates it,  especially at $T=398$ MeV. 
For this reason, we also consider a more involved fit function with three parameters, namely
\beq\label{eq:canon_08}
B(\theta_I) \simeq  {c \sin\big(w\theta_I\big) \over 1+\zeta\cos(w\theta_I) }\;,
\eeq
which is motivated by equations (\ref{eq:sin_fit_irrat}) and (\ref{eq:ff_for_Z}).
It is seen in Fig.~\ref{fig:qdr_Nt08_sin_tot_02} that this fit function provides an appropriate extrapolation
to the domain of real $\theta$. The fit results are presented in Table~\ref{tab:3fits}. 

\begin{figure}[tbh]
\vspace*{-23mm}
\hspace*{-7mm}\includegraphics[width=10cm]{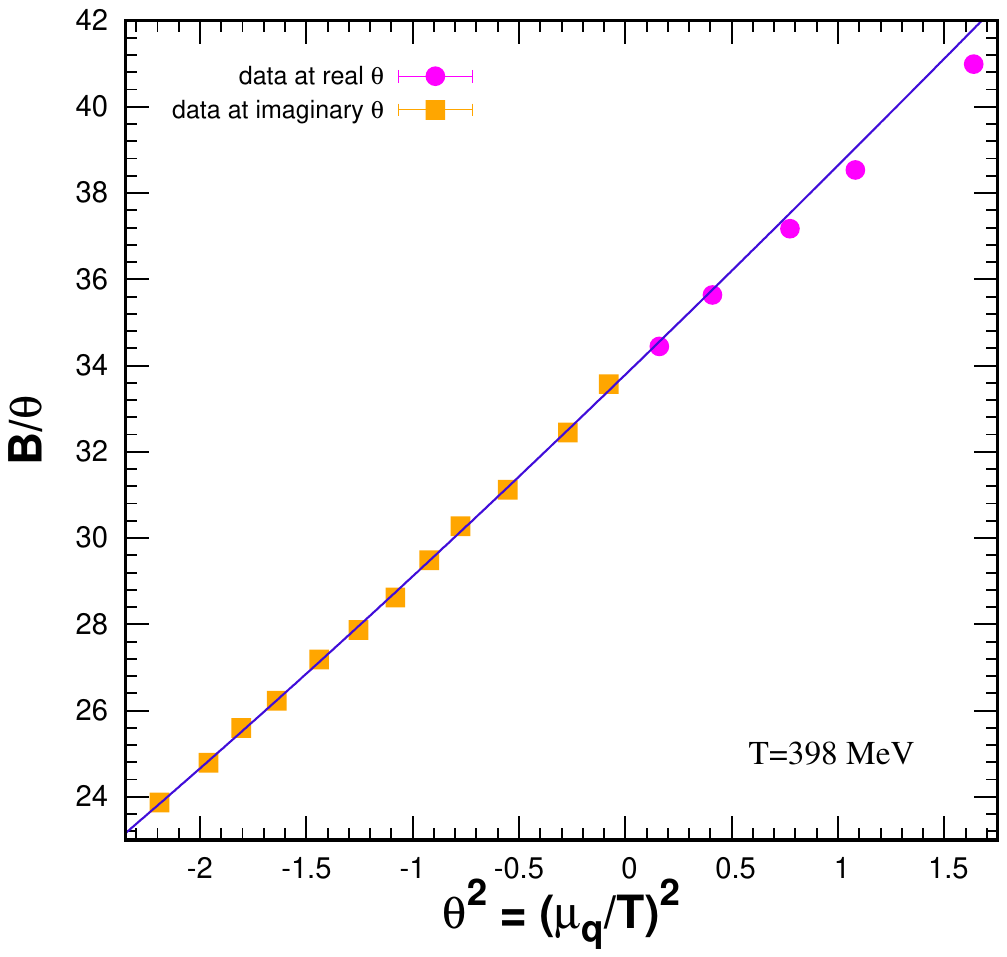}\hspace*{-10mm}
\includegraphics[width=10cm]{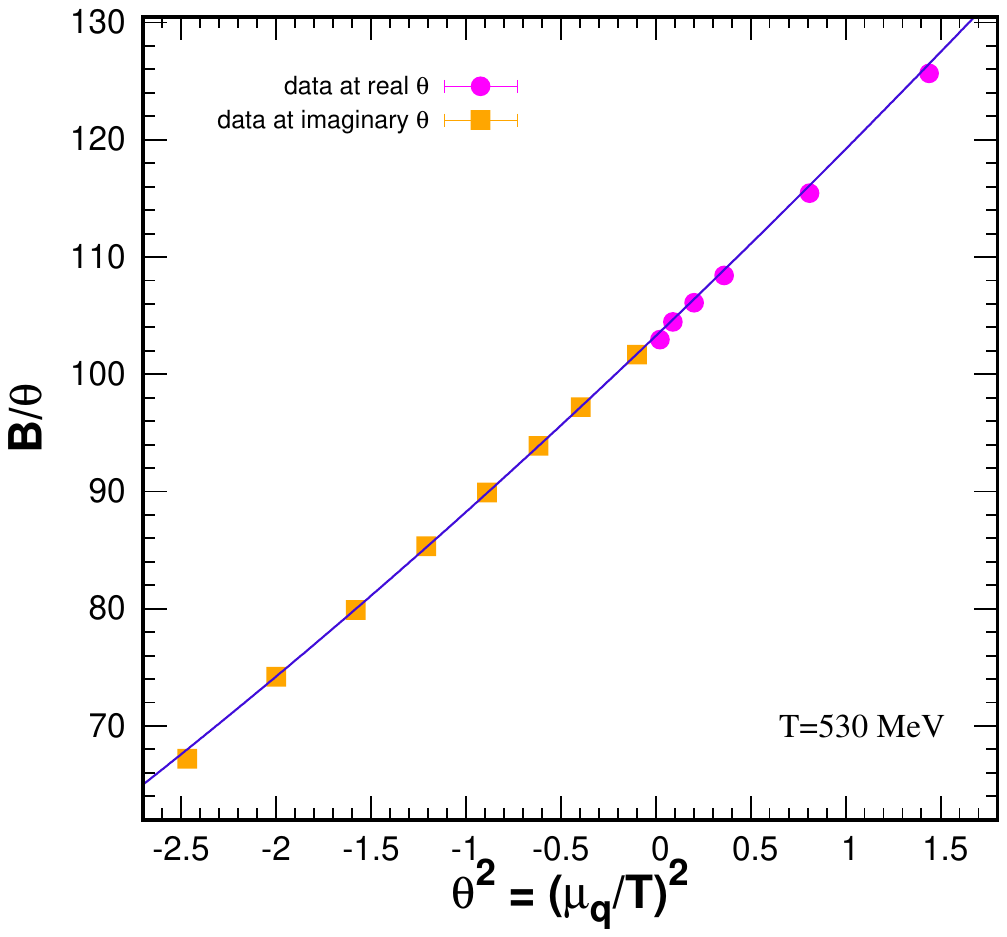}
\vspace*{-31mm}
\caption{The results of analytic continuation by formula
(\ref{eq:fitfun_w_eq_1}).}
\label{fig:extrap_sin_fit}
\end{figure}

It is seen in Table~\ref{tab:3fits} that the values of the parameter $w$ in the argument of the trigonometric functions is close to unity.
For this reason, we take $w=1$ 
and perform the fitting using the simplified formula
\beq\label{eq:fitfun_w_eq_1}
\tilde{B} \simeq  {c \sin\big(\theta_I\big) \over 1+\zeta\cos(\theta_I) }\;.
\eeq
The results are shown in Table~\ref{tab:3fits}. 
Goodness of fit is  characterized by $\chi^2/N_{dof} \lesssim 1$ and the fit parameters are well determined.
The analytic continuation using this formula agrees well with the data at real $\theta$, as seen in Fig.~\ref{fig:extrap_sin_fit}. 
Therefore, at high $T$ both the fit by a polynomial~(Eq.~\eqref{eq:pow_fit}) and by trigonometric functions~(e.g. Eq.~\eqref{eq:fitfun_w_eq_1})
can be used equally well for the analytic continuation to real values of $\theta$ in the range considered in this paper.

It is worth to note that the fit function (\ref{eq:fitfun_w_eq_1}) can be continued by analyticity 
to a $2\pi\imath$-periodic function of $\theta$  associated 
with the positive Polyakov-loop sector in the Roberge-Weiss approach.
The respective partition function provides 
an interesting toy model at $\ds {2c\over \zeta}=n\in Z\!\!\!Z$,
which possesses two high-order Lee-Yang zeroes. 
This can be an interesting subject of future studies.
%%%%%%%%%%%%%%%%%%%%%%%%%%%%%%% Table V new %%%%%%%%%%%%%%%%%%%%%%%%%%%%%%%
\begin{table}[hhh]
\bc
\begin{tabular}{|c|c|c|c|c|} \hline
~$T\;$(MeV)~ &  $\chi^2/N_{dof}$ & ~$c$~  &  ~$\zeta $~ & ~$w$~ \\ \hline
\multicolumn{5}{|c|}{Fit Eq.~(\ref{eq:sin_fit_irrat})} \\ \hline
 398 &  0.60  &  35.97(12)   &  0 &  0.9415(43)  \\
 530 &  2.9  &  107.45(23)   &  0 &  0.9627(28)  \\
\hline 
\multicolumn{5}{|c|}{Fit Eq.~(\ref{eq:canon_08})} \\ \hline
 398 &  0.64  &  35.78(32)   & 0.030(43) &  0.973(47)  \\
 530 &  0.79  &  106.42(30)   & 0.053(13) &  1.020(14)  \\
\hline 
\multicolumn{5}{|c|}{Fit Eq.~(\ref{eq:fitfun_w_eq_1})} \\ \hline
 398 &  0.60  &  35.601(84)   & 0.0536(39) &     1  \\
 530 &  0.95  &  106.426(98)   & 0.0345(15) &    1  \\
\hline
\end{tabular}
\ec
\caption{The results of the fits by the Eqs.~(\ref{eq:sin_fit_irrat}), (\ref{eq:canon_08}) and (\ref{eq:fitfun_w_eq_1}).
}
\label{tab:3fits}
\end{table}
%%%%%%%%%%%%%%%%%%%%%%%%%%%%%%%% end Table V %%%%%%%%%%%%%%%%%%%%%%%%%%%%%%%%%%%%%%%%%

%--------------------------------------------%%%%% FIGURE
%%%%% END FIGURE
%----------------------------------------------------------

\section{Conclusions}
\label{sec:conclusions}

We have studied the properties of quark number density $n_q$ at real and imaginary values of the quark chemical potential $\mu_q$ in lattice QC$_2$D over the temperature range 230 MeV $ \lesssim T \lesssim 530$~MeV. 
The analysis has been focused on the performance 
of various methods of extrapolation 
from imaginary to real chemical potential.
This analysis should be useful for the case of 
real QCD, where the lattice simulations are hindered 
by the sign problem at real chemical potentials 
but remain feasible at imaginary chemical potentials.
The analysis is  focused separately on temperatures below the Roberge-Weiss transition at $T_{\rm RW}$  and above.

At low temperatures ($T<T_{\rm RW}$), 
we first considered the analysis using model-independent expressions, such as the truncated trigonometric series expansion of the baryon number at imaginary chemical potentials~(see Eq.~(\ref{eq:Fourier_expansion})) and the expression Eq.~(\ref{eq:ff_for_Z}) based on the canonical formalism.
The former approach has shown only a limited success due to a small radius of convergence in the $\xi$ plane caused by a slow decrease of $a_n$ with increasing $n$. The latter approach also did not
demonstrate a large range of agreement with lattice data for the real chemical potential. In this case, the insufficient precision of our data at the imaginary chemical potential did not allow us to determine the high-order expansion coefficients, $Z_n$.

To proceed further, we then used fit functions based on two phenomenological models that have recently been proposed in the literature, the cluster expansion model~(CEM) and the rational fraction model~(RFM).
Both of these models make it possible to express all coefficients  
of the Fourier series for the quark density 
on the imaginary chemical potential segment $\ds -\;{\pi\over 2}\leq \theta_I 
\leq {\pi\over 2} (\theta_R=0)$ 
in terms of two parameters that are readily determined by fitting.
Moreover, the functions (\ref{CEM_summed}) and
(\ref{eq:RFM_summed_series_2}) obtained by summation
of these Fourier series for the CEM and the RFM, respectively,
 provide an analytic continuation of the quark density
to the real values of $\theta$.
We find that the CEM demonstrates a wide range of agreement with our lattice data at real $\theta$, while the RFM shows significant deviations from the data.

We further studied the canonical partition functions $Z_n$
predicted by CEM and RFM. 
We have found that the positivity of $Z_n$ 
is broken for both the CEM and the RFM for sufficiently high $n$, which is, most probably, due to unphysical singularities of the respective expressions.
From this we conclude that either corrections 
to the models, in particular at high quark densities, 
are required to recover positivity of $Z_n$, or that finite-volume effects are still substantial and not under control when $n$ is large.

We have shown that when only a limited number 
$N \leq N_{max}$ of the coefficients $Z_n\; (n<N)$ 
based on the CEM model is considered 
($N_{max}$  is a number such that for $n>N_{max}$ the sign of $Z_n$ either alternates or becomes negative), the extrapolation based on the canonical approach  agrees with the lattice data over a finite range $\ds 0\leq \theta_R\leq \theta^{(Z)}_R(N)$, where $\theta^{(Z)}_R(N)$ increases with~$N$.
We see that to extend the range of extrapolation based on the canonical formalism we need to determine the higher order coefficients $Z_n$ with high enough precision, either via direct fit to the lattice data at imaginary $\theta$ or by improving
the CEM at high densities. This should be the subject of a future work.

We also plan to extend our study to lower temperatures to cover, in particular, the vicinity of the 
transition to the quark-gluon plasma phase.
Finite volume effects should be also studied in future, and we do plan to simulate larger lattices with $N_s$ up to 56, in addition to the present value, $N_s=28$.
An increased statistics of the simulations will also allow us to compute $Z_n$ for higher values of $n$ using the direct fit approach. 

For high temperatures~($T>T_{\rm RW}$)
we have found that the quark density 
at imaginary quark chemical potential over the range 
$\ds 0<\theta_I<{\pi\over 2}$ can be equally well approximated either by a polynomial that can 
be associated with
the density of colored massless free fermions or by a $2\pi$-periodic trigonometric function (\ref{eq:fitfun_w_eq_1}).
The extrapolation to the real values 
of $\theta$ based on either of these fit-functions works 
equally well at least up to $\mu_q =600$~MeV. 

Our results have been  obtained for rather large pion mass. This implies that our conclusions should be checked in simulations with substantially smaller pion mass. Still,  our pion mass is in the range of the crossover (at zero quark density) and we expect that 
qualitatively our conclusions will stay intact. These expectations are also supported by agreement between lattice results for the quark number density and the chiral perturbation theory predictions found in Ref.~\cite{Astrakhantsev:2020tdl} at low temperature for equally heavy quarks ($m_\pi \approx 740 $ MeV).
We believe that the canonical approach to extrapolation from imaginary to real chemical potential can be 
applied to real QCD studies and we plan this work in future.

\acknowledgments{The authors are grateful to V.~Braguta and 
A.~Nikolaev for useful discussions and to V.~Vovchenko for his contribution in the early stages of this work.
The work was supported by the grant of the Russian Foundation for Basic Research No. 18-02-40130 mega and partially carried out within the state assignment of the Ministry of Science and Higher Education of Russia (Project No. FZNS-2024-0002). 
This work was partially supported by Grants-in-Aid for Scientific Research (Kakenhi), No. 21K03573.
Computer simulations were performed on the FEFU GPU cluster Vostok-1, the Central Linux Cluster of the 
NRC ”Kurchatov Institute” - IHEP (Protvino), the Linux Cluster of the NRC ”Kurchatov Institute” - ITEP (Moscow). In addition, we used computer resources 
of the federal collective usage center Complex for Simulation and Data Processing for Mega-science Facilities at NRC Kurchatov Institute,
http://ckp.nrcki.ru/.
A.B. is supported by the Carl Trygger Foundation Grant No. CTS 18:276. Nordita is supported in part by Nordforsk.}

\appendix

\section{}
\label{sec_App1}
We found that to determine $Z_n$ via the fit to Eq.~(\ref{eq:ff_for_Z})
the initial values of 
the fitting parameters 
must be very close to their 
final values in order to make the fitting algorithm properly convergent.
For example, taking for initial values of $Z_n$ their final values shifted by only 1\% 
may give rise to divergent fit. 
%%%
To deal with this problem, we use the following procedure.
First we find the Fourier coefficients $a_n$ 
by performing a fit to the function 
\beq\label{eq:Fourier_expansion_N}
 \tilde{B}^{(a)}_N(\theta_I)=\sum_{n=1}^{N_a} a_n \sin\left({2 n \theta_I}\right),  
 \qquad \theta_I \in \left[0, {\pi\over 2}\right] 
\eeq
with an appropriate value of $N_a$. Then 
$Z_n$ can be computed using a truncated version of
the relation 
\beq\label{eq:a_Z_identity}
\sum_{n=1}^\infty a_n \sin\left({2 n \theta_I}\right)  = {2 \sum_{n=1}^\infty n Z_n \sin(2n\theta_I) \over 1 + 2 \sum_{n=1}^\infty  Z_n \cos(2n\theta_I)}. 
\eeq

Multiplying both sides of Eq.~(\ref{eq:a_Z_identity})
by the denominator of the right hand side 
and employing trigonometric identities, we derive 
the following relations between $a_n$ and $Z_n$:
\beq\label{eq:aZ_sys_eq}
a_n=\sum_{m=1}^{\infty} W_{nm} Z_m \;,
\eeq
where
\beq\label{eq:Wmatrix}
{W}_{nm} = 2n \delta_{nm} - a_{n+m} + a_{|n-m|}\cdot \mbox{sign}(m-n) \;,
\eeq
where sign(0)=0 by definition. 
As $Z_m$ rapidly decrease with
increasing $m$, we can neglect the terms in the rhs of eq.~(\ref{eq:aZ_sys_eq})
starting from some $m=N+1$. 
Thus, we arrive at 
the linear system of $N$ equations
for determination of $Z_m$ in terms of $a_n$.
This system can be cast in the matrix form
where $\mathbf{a}$ is the column vector of the coefficients $a_n$ and
elements of the square matrix $\mathbf{W}$ 
have the form~(\ref{eq:Wmatrix}).
To find $N$ coefficients $Z_n$ we need $2N$ coefficients $a_{n}$
because the $N\times N$ matrix $\mathbf{W}$ involves $a_n$ for $n=1,2,...,2N$. 
This is used in case of models CEM and RFM when all $a_n$ are known. To find initial values of the parameters $Z_n$ when fitting to the function (\ref{eq:ff_for_Z}), it is sufficient to use several $a_n$ and take the remaining $a_n$ equal to zero.

\section{}
\label{sec_App2}
%%%%%%%%%%%%%%%%%%%%%  BEGIN TABLE VII ###########################
\begin{table}[thb]
\begin{tabular}{|c|c|c|c|c|c|c|c|c|c|} \hline
\multicolumn{5}{|c|}{Fit eq.~(\ref{eq:ff_for_Z})} & \multicolumn{5}{|c|}{Fit eq.~(\ref{eq:Fourier_expansion})} \\
    \hline
~~$N$~~& $~n~$ & $Z_n$ & $\chi^2/N_{dof}$ &$p$-value& ~~$N$~~&
$~n~$   & $a_n$   &  $\chi^2/N_{dof}$  &$p$-value  \\ 
\hline
2& 1 &---& 47 & $0$  & 2 & 1 & $1.928(29) $ & 1.66 & $0.069$            \\
 & 2 &---&     &     & & 2 & $-0.164(24) $ &     &  \\ \hline 
 & 1 &  $0.6694(38)$ &         &      &   & 1 & $1.917(30) $ &      &      \\
3& 2 & $0.2506(39) $`&  $2.21$ & $0.012$ & 3 & 2 &$-0.128(33) $ & 1.57 &  0.10 \\
 & 3 & $0.0510(21) $ &      &            &   & 3 & $-0.036(24) $&      & \\
\hline
 &  1 &  $0.6709(40)$ &     &      &   & 1& $ 1.918(30) $ &      &     \\
4& 2 &  $0.2553(45)$ & 1.27 & 0.24 & 4 & 2& $-0.127(33) $ & 1.71 & 0.072    \\
 & 3 &  $0.0580(33)$ & &           &   & 3& $-0.046(36) $ & &    \\
 & 4 &  $0.0049(16)$ & &           &   & 4& $ 0.011(26) $ & &    \\
\hline 
 & 1 &  $0.6708(40)$&       &      &   & 1& $ 1.918(30) $ &   &       \\
 & 2 & $0.2551(45)$ &       &      &   & 2& $-0.139(34) $ &   &\\
5& 3  & $0.0582(33)$&  1.37 & 0.19 & 5 & 3& $-0.025(37) $ &1.09 & 0.36 \\
 & 4  &  $0.0057(21)$&      &      &   & 4& $-0.048(36) $ &   & \\
 &  5  &  $0.0007(11)$&     &      &   & 5& $ 0.067(26) $ &   &    \\ 
 \hline 
 & 1 &  $0.6702(40)$&       &      &   & 1& $ 1.922(30) $ &   &       \\
 & 2 & $0.2536(46)$ &       &      &   & 2& $-0.142(34) $ &   &\\
6& 3  & $0.0567(34)$&  1.19 & 0.30 & 6 & 3& $-0.025(37) $ &1.10 & 0.36 \\
 & 4  &  $0.0060(21)$&      &      &   & 4& $-0.058(37) $ &   & \\
 &  5  &  $0.0028(17)$&     &      &   & 5& $ 0.093(39) $ &   &    \\
 &  6  &  $0.0017(10)$&     &      &   & 6& $-0.024(25) $ &   &    \\
  \hline 
& 1 &  $0.6697(40)$&       &      &   & &   &   &       \\
 & 2 & $0.2527(46)$ &       &      &   & &  &   &\\
& 3  & $0.0562(34)$&        &      &   & &  & &  \\
7& 4  &  $0.0059(21)$& 1.21 & 0.29 &   & &  &   & \\
 &  5  &  $0.0031(17)$&     &      &   & &  &   &    \\
 &  6  &  $0.0026(13)$&     &      &   & &  &   &    \\
  &  7  &  $0.0008(8)$&     &      &   & &  &   &    \\
\hline  
\end{tabular}
\caption{The coefficients $Z_n$ for the fit function (\ref{eq:ff_for_Z}) 
and $a_n$ for the fit 
function (\ref{eq:Fourier_expansion})   determined from the fits 
to our data  over the range 
$\ds 0\leq \theta_I \leq  {\pi\over 2} $  at $T=227$~MeV.
We compare results for four values of $N$ in 
Eqs.~(\ref{eq:ff_for_Z}) and ~(\ref{eq:Fourier_expansion}).
Parameters characterizing goodness of fit are also shown. }
\label{tab:direct_fit_14}
\end{table}
%%%%%%%%%%%%%%%%%%%%%    END TABLE VII  ############################
%%%%%%%%%%%%%%%%%%%%%  BEGIN TABLE VIII  ############################
\begin{table}[t]
\begin{tabular}{|c|c|c|c|c|c|c|c|c|c|} \hline
\multicolumn{5}{|c|}{Fit eq.~(\ref{eq:ff_for_Z})} & \multicolumn{5}{|c|}{Fit eq.~(\ref{eq:Fourier_expansion})} \\
    \hline
$~N~$ & $~n~$ &  ~$Z_n$~ & $\chi^2/N_{dof}$&$p-$value & $~N~$ & $~n~$ &  ~$a_n$~ & $\chi^2/N_{dof}$  & $p-$value \\ \hline
      &  1    &0.860322(58)&       &        &    & 1 & 5.040(18)&      &  \\
      &  2    &0.551440(56)&        &        &    & 2 & -1.173(18)&     &    \\
      &  3    &0.26742(17)&       &        &  5 & 3 & 0.423(22)& 0.49 & 0.91 \\   
      &  4    &0.09997(39)&        &        &    & 4 & -0.164(19)&     &    \\ 
   8  &  5    &0.02938(42)&  0.32  &  0.97&    & 5 & 0.047(11)&     &    \\ 
      &  6    &0.00691(29)&        &        &    &   &           &     &    \\
      &  7    &0.00127(13)&        &        &    &   &           &     &    \\
      &  8    &0.000143(28))&        &        &    &   &           &     &    \\
\hline 
%%%%%%%%%%%%%%%%%%%%%%%%%%%%%%%%%%%%%%%%%%%%%%%%%%%%%%%%%%%%%%%%%%%%%%%%%%%%%%   
      &  1    &0.860321(34)&       &  &    & 1 &5.040 (18)&      &  \\
      &  2    &0.551446(39) &       &  &    & 2 &-1.175(19) &      &  \\ 
      &  3    &0.267419(61)&       &  &  6 & 3 &0.430(22)& 0.26 & 0.99  \\
      &  4    &0.099972(44) &       &  &    & 4 &-0.185(25)&      &  \\
   9   &  5    &0.02942(13)& 0.48 & 0.85& & 5 &0.080(23) &      &  \\
      &  6    &0.00696(24) &       &  &    & 6 &-0.024(13)&      &  \\
      &  7    &0.00130(13) &       &  &    &   &         &      &  \\
      &  8    &0.000148(82) &       &  &    &   &         &      &  \\
      &  9    &0.000003(23) &       &  &    &   &         &      &  \\
\hline
\end{tabular}
\caption{
Same as in Table \ref{tab:direct_fit_14} but for $T=265$~MeV and different values of $N$ in Eqs.~(\ref{eq:Fourier_expansion}) and (\ref{eq:ff_for_Z}).
}
\label{tab:direct_fit_12}
\end{table}
%%%%%%%%%%%%%%%%%%%%%    END TABLE VIII  
\end{document}